\documentclass[pra,twocolumn,tbtags,superscriptaddress,floatfix,showpacs]{revtex4}
\usepackage{amssymb}
\usepackage{graphicx,amsmath}
\usepackage{bm}
\usepackage{times}
\begin{document}

\title{Non-Hermitian Hamiltonian approach to the microwave
transmission through one- dimensional  qubit chain}

\begin{abstract}
We investigate the propagation of microwave photons in a
one-dimensional open waveguide interacting with a number of
artificial atoms (qubits). Within the formalism of projection
operators and non-Hermitian Hamiltonian approach we develop a
one-photon approximation scheme for the calculation of the
transmission and reflection factors of the microwave signal in a
waveguide which contains an arbitrary number \emph{N} of
non-interacting qubits. We considered in detail the resonances and
photon mediated entanglement for two and three qubits in a chain.
We showed that in non Markovian case the resonance widths, which
define the decay rates of the entangled state, can be much smaller
than the decay width of individual qubit. It is also shown that
for identical qubits in the long wavelength limit a coherent
superradiant state is formed with the width being equal to the sum
of the widths of spontaneous transitions of \emph{N} individual
qubits. The results obtained in the paper are of general nature
and can be applied to any type of qubits. The specific properties
of the qubit are only encoded in the two parameters: the qubit
energy $\Omega$ and the rate of spontaneous emission $\Gamma$.
\end{abstract}


\date{\today }
\author{Ya. S. Greenberg}\email{yakovgreenberg@yahoo.com}
\affiliation{Novosibirsk State Technical University, Novosibirsk,
Russia}
\author{A. A. Shtygashev}
\affiliation{Novosibirsk State Technical University, Novosibirsk,
Russia}


 \maketitle


\section{Introduction}
One-dimensional (1D) waveguide-quantum electrodynamics (QED)
systems are emerging as promising candidates for quantum
information processing motivated by tremendous experimental
progress in a wide variety of solid state systems with imbedded
artificially designed atoms- qubits \cite{Buluta2011, You2011,
Buluta2009, Fang15}. Confining the microwave field in reduced
dimensions such as 1D waveguide and taking account of the enormous
dipole moment of artificial atom the photon qubit interaction can
be strongly enhanced as compared with open 3D space \cite{Blais04,
Wallraff04}. In recent years one of the basic type of these
physical systems has been realized in solid state setups where
qubits were on chip coupled to microwave cavities \cite{Schoe08}.
An important advantage of these systems is that the qubits can be
placed within the photon field confined in a microwave cavity at
fixed predetermined positions at separations on the order of
relevant wavelength. Moreover, unlike the real atoms, qubits are
intrinsically not identical due to technological scattering of
their parameters.  It is also important that the excitation energy
of every qubit in a chain can easily be adjusted by external
circuit.

The experimental investigation of these systems is based on the
measurements of the transmitted and reflected signals with their
properties being dependent on the quantum states of every qubit in
a waveguide. Up to now there are known only several experiments
with a single superconducting qubits in 1D open space
\cite{Ast10a, Ast10b, Abdul10, Abdul11,Hoi11, Hoi12, Hoi13a,
Hoi13b} and one experiment with two transmon- type qubits in a
waveguide. \cite{Loo13}.

For solid-state quantum information processing it is interesting
to study 1D waveguide systems having more than just one qubit. A
key point here is whether the multi-qubit system could display a
long lived entanglement necessary for implementation of quantum
algorithms. The entanglement is also necessary for quantum error
correction which requires at least three qubits in the
chain\cite{DiCarlo00}.

Such multi-qubit systems exhibit, in general, non Markovian
behavior: the interaction between qubits is not instantaneous,
hence, the retardation effects have to be included. The
manifestation of these effects is that the resonances (energies
and their widths) of the qubit system become dependent on the
frequency of incident photon. In this case a master equation for
the density matrix $\rho$ of the qubits cannot be written in
Lindblad form. A Markovian approximation corresponds to long
wavelength limit, $kd<<1$, where $k$ is photon wave vector, $d$ is
a distance between neighbor qubits. In this case we may neglect
the retardation effects and assume that the qubits interact
instantaneously.

Recent experiments with superconducting qubits showed that the
photon mediated interaction between distant qubits can lead to the
creation of two-qubit \cite{Roch14, Loo13} and multi-qubit
entanglement\cite{Jerger11, Jerger12, Macha14}.

Theoretical calculations of microwaves transmission in 1D open
waveguide with a qubit placed inside are being performed in a
configuration space \cite{Shen05a, Shen05b, Zheng13, Fang14,
Shen09} or by the input- output formalism \cite{Fan10,
Lalumiere13}. These methods are physically sound but they become
very cumbersome if we try to find solutions for several or more
qubits in a waveguide. While the transmission for a single two
level atom in 1D open waveguide has long been known \cite{Shen05a,
Shen05b}, the analytical expressions for the transmissions for two
qubits and for symmetrical arrangement of three identical qubits
have been published quite recently \cite{Zheng13, Fang14}.

For N identical equally spaced qubits the transmission can be
found analytically with the help of the method borrowed from the
physics of crystals with translational symmetry \cite{Tsoi08}.
However, in general, for N non identical randomly spaced qubits a
simple analytical procedure does not exist.

In the present paper we propose a matrix formalism for the study
of a one- photon transport in 1D open waveguide filled, in
general, with a number of not identical and arbitrary spaced
artificial atoms. Similar idea has been suggested for the study of
photon transport in the coupled resonator optical waveguides
\cite{Xu00}. Our approach is based on the projection operators
formalism and the method of the effective non- Hermitian
Hamiltonian which are the powerful tools to deal with a
Lippman-Shwinger scattering problem. It is different from usual
resolvent method of solving Lippmann-Schwinger equation which was
demonstrated for two qubits in \cite{Diaz15}.

The method we use here has originally been developed for the
description of nuclear reactions \cite{Feshbach58, Sokolov92} with
many later applications for different open mesoscopic systems
ranging from universal conductance fluctuations \cite{Sor09} to
electron transport through 1D solid state nanostructures
\cite{Celardo10, Greenberg13}(see review paper \cite{Auerbach11}
and references therein).

For general $N$ qubit case our technique allows us to easily
include the cases of non identical qubits and/or with unequal
spacing when the translation symmetry is absent. It is very
important for artificial atoms with inevitable technological
spreading of parameters with the possible individual tuning of
qubit resonance energies. Additionally, a direct exchange
interaction  between nearest neighbor qubits can be easily
incorporated into the scheme of this technique. The influence of
this interaction between two and $N$ superconducting flux qubits
on the photon transmission and entanglement has been studied in
the papers\cite{Ilichev2010, Temchenko2011, Zippilli2015}.

With the aid of our technique we study in detail the one- photon
microwave transport for one, two and three qubits imbedded in a
waveguide. We considered in detail the resonances and photon
mediated entanglement for two and three qubits in a chain. We show
that for $N$ identical qubits in the long-wavelength limit a
coherent superradiance state is formed with the width being equal
to the sum of the widths of spontaneous transitions of \emph{N}
individual qubits.

The paper is organized as follows. In the Section II we define the
model Hamiltonian of $N$ noninteracting qubits imbedded in a
microwave resonator. In Section III we describe in detail a
projection formalism and effective non-Hermitian Hamiltonian
approach in application to the photon transport in 1D waveguide.
The application of the model Hamiltonian to the derivation of the
general expressions for the transmission and reflection
coefficients for $N$ qubits in a waveguide is performed in the
Section IV. The Section V is devoted to a detailed investigation
of the microwave transport for one, two, and three qubits in a
waveguide. In this section we give not only the analytical
expressions for the transmission and reflection factors for two
and three qubits in general case, but we investigate in detail the
energy spectrum of these systems and their resonances in non
Markovian case, which is automatically included in our theory,
since the quantity $kd$ explicitly enters the analytical
expressions. We also study a photon mediated entanglement for two
and three qubit systems. The results of this section are important
for three qubits experiments, since to our knowledge there are no
1D open space experiments with three qubits in a waveguide. In the
conclusion to this section we briefly analyze the general case of
$N$ qubits.

\section{The model Hamiltonian}\label{2}
We consider a microwave 1D waveguide resonator with \emph{N}
qubits imbedded at the fixed positions $x_i$. The Hamiltonian of
the system reads:
\begin{equation}\label{H}
    H = {H_{ph}} + {H_{qb}} + {H_{{\mathop{\rm int}} }}
\end{equation}
where
\begin{equation}\label{H_ph}
    {H_{ph}} = \sum\limits_k {\hbar {\omega _k}a_k^ + {a_k}}
\end{equation}
is the Hamiltonian of photon field,
\begin{equation}\label{H_qb}
    {H_{qb}} = \sum\nolimits_{i=1}^N {H_{qb}^i}
\end{equation}
is the Hamiltonian of \emph{N} noninteracting qubits, where
\begin{equation}\label{H_1qb}
    H_{qb}^i = \frac{1}{2}\hbar {\Omega _i}\sigma _z^{(i)}
\end{equation}
is the Hamiltonian of the individual $i$-th qubit with the
excitation frequency $\Omega_i$.

The interaction of the qubit chain with the photon field is given
by Hamiltonian:

\begin{equation}\label{H_int}
    {H_{{\mathop{\rm int}} }} = \sum_k\sum\nolimits_{i=1}^N {{\lambda _i}}
    (a_k^ + {e^{ - ik{x_i}}} + {a_k}{e^{ik{x_i}}})\sigma _x^{(i)}
\end{equation}
where $\lambda_i$ is the qubit-photon interaction strength, $x_i$
are the qubit positions relative to the waveguide center, $x_0=0$.

\section{Projection formalism and effective non-Hermitian
Hamiltonian}\label{3}
\subsubsection{Projection operators formalism from the formal point of view}

As the projection operators formalism and effective non-Hermitian
Hamiltonian approach are not common in the field of quantum
optics, here we briefly describe the essence of this method
omitting its rigorous justification which can be found in the
corresponding literature (see review paper \cite{Auerbach11} and
references therein).

It is always possible to formally subdivide the Hilbert space of a
quantum system with the Hermitian Hamiltonian $H$ into two
arbitrarily selected orthogonal projectors, $P$ and $Q$, which
satisfy the properties of completeness

\begin{equation}\label{compl}
    1=P+Q
\end{equation}
and orthogonality
\begin{equation}\label{ort}
    PQ=QP=0
\end{equation}
From (\ref{compl}) and (\ref{ort}) it also follows:
\begin{equation}\label{pp}
     PP=P, \ QQ=Q
\end{equation}

 With the help of the completeness (\ref{compl}) we can divide
 the solution of the stationary Schr\"{o}dinger equation,
\begin{equation}\label{Schrod}
    H\Psi=E\Psi
\end{equation}
in two parts,
\begin{equation}\label{decomp}
    \Psi\equiv P\Psi+Q\Psi\equiv \Psi_P+\Psi_Q
\end{equation}
and rewrite (\ref{Schrod}) in the following form:
\begin{equation}\label{Schrod1}
 (P+Q)H(P+Q)(\Psi_P+\Psi_Q)=E(\Psi_P+\Psi_Q)
\end{equation}
Since, in virtue of (\ref{ort}), $P\Psi_Q$=0, $Q\Psi_P=0$ we
rewrite (\ref{Schrod1}) as follows
\begin{multline}\label{Schrod2}
    (H_{PP}+H_{QP})\Psi_P+(H_{QQ}+H_{PQ})\Psi_Q=E(\Psi_P+\Psi_Q)
\end{multline}
where
\[ H_{QQ}=QHQ, H_{PP}=PHP, H_{QP}=QHP, H_{PQ}=PHQ
 \]
The equation (\ref{Schrod2}) is equivalent to the Schr\"{o}dinger
equation (\ref{Schrod}).

Multiplying (\ref{Schrod2}) from the left by projectors $P$ and
$Q$ we obtain two coupled equations for $\Psi_P$ and $\Psi_Q$:
\begin{equation}\label{psiP}
    (H_{PP}-E)\Psi_P=-H_{PQ}\Psi_Q
\end{equation}
\begin{equation}\label{psiQ}
    (H_{QQ}-E)\Psi_Q=-H_{QP}\Psi_P
\end{equation}


If we eliminate from (\ref{psiP}) or (\ref{psiQ}) one subspace of
states, we can obtain an equation for a part of the wave function
(\ref{decomp}).

For example, if we eliminate from (\ref{psiP}) the $P$-subspace,
the equation for the wave function in $Q$- subspace takes the form
\begin{equation}\label{H_Q}
    H_{eff}(E)\Psi_Q=E\Psi_Q
\end{equation}
where the \emph{energy dependent} effective Hamiltonian
\begin{equation}\label{Heff}
    {H_{eff}(E)} = {H_{QQ}} + {H_{QP}}\frac{1}{{E - {H_{PP}} }}{H_{PQ}}
\end{equation}

projects Hilbert space on the $Q$ subspace.

The second term in (\ref{Heff}) describes multiple excursions to
the class $P$ with return to the class $Q$.

It should be noted that while the energy $E$ in (\ref{H_Q}) is the
same as in Schr\"{o}dinger equation (\ref{Schrod}), the equation
(\ref{H_Q}) is not equivalent to (\ref{Schrod}): the effective
Hamiltonian (\ref{Heff}) is energy dependent , so that the
eigenvalue $E$ enters this equation in a complex way, and
wavefunction $\Psi_Q$ is not egenfunction for $E$, however it can
be written as a linear superposition of the state vectors from
$Q$- subspace.

\subsubsection{Application to the scattering problem}\label{3_2}

Keeping in mind the scattering problem we assume that $Q$ subspace
consists of discrete states, and $P$ subspace consists of the
states from continuum, however, it may also contain the discrete
states. We also assume that Hamiltonian $H_{PP}$ is diagonal in
subspace P. In order to avoid the singularities emerging when
$H_{PP}$ has eigenvalues at real energy $E$, it has to be
considered as a limiting value from the upper half of the complex
energy plane, $E^{+}=E+i\varepsilon$. With this rule, those states
of subspace Q which will turn out to be coupled to the states in
subspace P will acquire the outgoing waves and become unstable.
Then, for this scattering problem the effective Hamiltonian
(\ref{Heff}) becomes non Hermitian and  has to be written as
follows:
\begin{equation}\label{Heff1}
    {H_{eff}(E)} = {H_{QQ}} + {H_{QP}}\frac{1}{{E - {H_{PP}+i\varepsilon} }}{H_{PQ}}
\end{equation}
In this case the equation (\ref{H_Q}) defines the resonance
energies of the Q- system which lie in the low half of the complex
energy plane, $\rm{E}=\rm{\widetilde{E}}-i\hbar\widetilde{\Gamma}$
and are given by the roots of the equation
\begin{equation}\label{2qb2}
    \det \left( {{\rm E} - {H_{eff}}} \right) = 0
\end{equation}
The imaginary part $\widetilde{\Gamma}$ of the resonances
describes the decay of $Q$- states due to their interaction with
$P$- states.

In the framework of projection formalism we can find from
(\ref{psiP}), (\ref{psiQ}) the wavefunction of the whole system
$\Psi$, a solution of Shr\"{o}dinger equation (\ref{Schrod}), in
terms of the operator which acts on the initial state,
$|in\rangle$, which contains continuum variables and satisfies the
equation $H_{PP}|in \rangle =E |in \rangle$, where $E$ is the same
as in (\ref{Schrod}). Then, the formal solution of (\ref{psiP})
can be expressed in the following form
\begin{equation}\label{psiP1}
    \Psi_P=|in\rangle+\frac{1}{E-H_{PP}+i\varepsilon}H_{PQ}\Psi_Q
\end{equation}
Substituting this expression in r.h.s. of (\ref{psiQ}) we obtain
\begin{equation}\label{psiQ1}
    \Psi_Q=\frac{1}{E-H_{eff}}H_{QP}|in\rangle
\end{equation}
where $H_{eff}$ is given by its non Hermitian form (\ref{Heff1}).
As a final step, we substitute $\Psi_Q$ from (\ref{psiQ1}) into
r.h.s. of (\ref{psiP1}) and combine these two equations to obtain
the expression for the state vector of the Shr\"{o}dinger
wavefunction $\Psi$ \cite{Rotter01}

\begin{eqnarray}\label{Ph1}
    |{\Psi}\rangle  = \left| {in}\right\rangle
   + \frac{1}{{E - {H_{eff}}}}{H_{QP}}\left| {in} \right\rangle \nonumber\\
   +\frac{1}{{E - {H_{PP}} + i\varepsilon }} H_{PQ}\frac{1}{{E - {H_{eff}}}}{H_{QP}}\left| {in} \right\rangle\
\end{eqnarray}
In fact, this expression is nothing more than a decomposition
(\ref{decomp}). The last term in (\ref{Ph1}) is the part of
$\Psi_P$, which describes to all orders of $H_{QP}$ the evolution
of initial state $|in\rangle$ under the interaction between $P$
and $Q$ subspaces.

\subsubsection{One photon scattering}

In the one photon approximation there are two possibilities:
either one photon is in the waveguide in the state
$|1_k\rangle\equiv|k\rangle$ and all qubits are in their ground
states $|g_i\rangle$ with a corresponding state vector
$|g_1,g_2.........g_N,k\rangle$, or no photons in the waveguide,
$|0_k\rangle\equiv|0\rangle$, with $i$-th qubit being excited and
$N-1$ qubits being in their ground states. In this case the system
is described by $N$ vectors of the type
$|g_1,..g_{i-1},e_i,g_{i+1},...g_N,0\rangle$.

In order to simplify the notations we will use throughout the
paper the following concise forms for state vectors:
\begin{equation}\label{ket}
    |k\rangle\equiv|g_1,g_2.........g_N,k\rangle
\end{equation}
\begin{equation}\label{net}
    |n\rangle\equiv|g_1,..g_{n-1},e_n,g_{n+1},...g_N,0\rangle
\end{equation}

with the orthogonality relations
\[\langle n|m\rangle=\delta_{nm}\]
\[\langle n|k\rangle=0\]
\[\left\langle {k|k'} \right\rangle  = \frac{{2\pi
}}{L}\delta (k - k'),\] where $L$ is the waveguide length.

In these notations the initial state is just the state (\ref{ket})
$(|in\rangle\equiv|k\rangle)$ with the energy
\begin{equation}\label{en_in}
    E \equiv E_k= \hbar \omega_k  - \frac{\hbar }{2}\sum\limits_{i = 1}^N {{\Omega _i}}
\end{equation}
where $\omega_k$ is the frequency of incident photon.

 Hence we take the projection operators as follows:
\begin{equation}\label{P}
    P = \sum\nolimits_k {\left|k \right\rangle
    \left\langle k \right|}=\frac{L}{{2\pi }}\int\limits_{ - \infty }^{ + \infty }
    {dk\left| {k} \right\rangle \left\langle {k} \right|}
\end{equation}
\begin{equation}\label{Q}
    Q = \sum\limits_{n = 1}^N {\left| {n} \right\rangle \left\langle {n} \right|}
\end{equation}

 Then the matrix
elements of the effective Hamiltonian (\ref{Heff1}) in subspace Q
can be written as
\begin{equation}\label{H1}
    \left\langle {m} \right|{H_{eff}}\left| {n} \right\rangle  =
\left\langle {m} \right|H\left| {n} \right\rangle  +\nonumber
\end{equation}
\begin{equation}\label{H2}
    \frac{L}{2\pi}
    \int\limits_{ - \infty }^{ + \infty } {d{q}\frac{{\left\langle {m} \right|{H_{QP}}
    \left| {q} \right\rangle \left\langle {q} \right|{H_{PQ}}\left| {n}
    \right\rangle }}{{E_k - {E_q} + i\varepsilon }}}
\end{equation}

In the basis of Q- subspace vectors the full wavefunction
(\ref{Ph1}) can be written as
\begin{equation}\label{Ph2}
\begin{array}{l}
\left| \Psi  \right\rangle  = \left| {k} \right\rangle  +
\sum\limits_{n,m=1}^N {\left| n \right\rangle }
R_{n,m}\left\langle m \right|{H_{QP}}\left| {k}
\right\rangle  + \\
\frac{L}{{2\pi }}\sum\limits_{n,m=1}^N {\int\limits_{}^{} {dq}
\frac{{\left| {q} \right\rangle }}{{E_k - {E_q} + i\varepsilon
}}\left\langle {q} \right|{H_{PQ}}\left| n \right\rangle
R_{n,m}\left\langle m \right|{H_{QP}}\left| {k} \right\rangle }
\end{array}
\end{equation}

where $R_{m,n}$ is the matrix inverse of the matrix $\langle
m(E-H_{eff})|n\rangle$:
\begin{equation}\label{Rmn}
    R_{m,n}=\langle m|\frac{1}{E_k - H_{eff}}|n\rangle
\end{equation}

The second term in (\ref{Ph2}) is the wavefunction $\Psi_Q$ of a
qubit system modified by its interaction with a photon field:

\begin{equation}\label{QbWF}
    \Psi_Q=\sum\limits_{n,m=1}^N {\left| n \right\rangle }
R_{n,m}\left\langle m \right|{H_{QP}}\left| {k} \right\rangle
\end{equation}
In more general context the expression (\ref{QbWF}) describes the
entanglement between qubits due to their interaction with a photon
field.

 From (\ref{QbWF}) we can also find the probability for the $n$- th
qubit to be in excited state:
\begin{equation}\label{prob}
\left\langle {n|\Psi_Q } \right\rangle  = \sum\limits_{m=1}^N
{R_{n,m}^{}\left\langle m \right|{H_{QP}}\left| {k} \right\rangle}
\end{equation}

The photon wavefunction in configuration space is obtained by
multiplying (\ref{Ph2}) from the left by the vector $\langle
x|\equiv\langle x, g_N, g_{N-1},.....g_1|$:
\begin{equation}\label{Ph3}
\begin{array}{l}
\Psi_N(x)=\left\langle {x|\Psi } \right\rangle  = {e^{ikx}} + \\
\frac{L}{{2\pi }}\sum\limits_{n,m=1}^N {\int\limits_{}^{} {dq}
\frac{{{e^{iqx}}}}{{E_k- {E_q} + i\varepsilon }}\left\langle {q}
\right|{H_{PQ}}\left| n \right\rangle {R_{n,m}}\left\langle m
\right|{H_{QP}}\left| {k} \right\rangle }
\end{array}
\end{equation}
where we have used the definitions $\left\langle {x|k}
\right\rangle  = {e^{ikx}}$  and  $\left\langle {x|n}
\right\rangle  = 0$.

The wavefunction (\ref{Ph3}) is a superposition of the incident
wave and the wave which results from the virtual transitions
between qubits and photon field in the resonator. We will see
below that this superposition leads to the destructive
interference when the frequency of incident photon is equal to the
excitation frequency of any qubit. In this case the transmitted
signal outside the qubit array is equal to zero.

\section{The application of projection formalism to the model hamiltonian}\label{4}
Here we apply the model Hamiltonian (\ref{H}) from the Section
\ref{2} to the general expressions found in Section \ref{3}. First
we calculate the matrix elements of the effective Hamiltonian
(\ref{H2}). The first term in rhs of (\ref{H1}) reads:
\begin{equation}\label{Heps}
     \left\langle {m} \right|H\left| {n}
     \right\rangle=\varepsilon_m\delta_{m,n}
\end{equation}
where
\begin{equation}\label{Hij1}
    \varepsilon_m
    =\frac{1}{2}\hbar\left(\Omega_m-\sum\limits_{n\neq
    m}^{N}\Omega_n\right)
\end{equation}
It is also not difficult to calculate the matrix elements in rhs
of equation (\ref{H1}):
\begin{equation}\label{AA}
    \langle m|H_{QP}|k\rangle=\lambda_m \exp({ikx_m})
\end{equation}
Then, the the second term in rhs of (\ref{H1}) can be written as
\begin{equation}\label{Hef}
   \left(\frac{\lambda_m\lambda_n L}{2\pi}\right) J(x_m,x_n)
\end{equation}
where
\begin{equation}\label{Jij}
    J(x_m,x_n)=\int\limits_{ - \infty }^{ + \infty }{d{q}\frac{\exp(iq(x_m-x_n))}{{E_k - {E_q} +
i\varepsilon }}}
\end{equation}
It is shown in the Appendix that
\begin{equation}\label{Jij1}
    {J(x_m,x_n)}  =  - \frac{2i\pi}{\hbar v_g} {e^{ik|d_{mn}|}}
\end{equation}
where $d_{mn}=x_m-x_n$, and $k$ is related to the physical
frequency $\omega$ of incident photon, $k=\omega/v_g$, where
$v_g$ is the group velocity of the photon wave in a waveguide.

Finally, the effective Hamiltonian (\ref{H1}) can be written as
follows
\begin{multline}\label{Hij2}
    \left\langle {m} \right|{H_{eff}}\left| {n} \right\rangle  =
\varepsilon_m\delta_{m,n}
   -i\hbar(\Gamma_m\Gamma_n)^{1/2}e^{ik|d_{mn}|}
\end{multline}
where we define the halfwidth of spontaneous emission
\begin{equation}\label{SE}
    \Gamma_m=\frac{L\lambda_m^2}{\hbar^2v_g}
\end{equation}

Throughout the paper we will use $\Gamma$ for the \emph{halfwidth}
of resonance line.

The photon wavefunction (\ref{Ph3}) for our model follows from
(\ref{AA}) and (\ref{Jij1}):

\begin{equation}\label{Ph4}
   \Psi_N(x)=e^{ikx}-i\hbar\sum\limits_{m,n = 1}^N
     (\Gamma_m\Gamma_n)^{1/2}{e^{ikx_m}R_{m,n}e^{ik|x-x_n|}}
\end{equation}
where the matrix $R_{m,n}$ is defined in (ref{Rmn}).

Finally, for our model we write down the qubits' wavefunction
$\Psi_Q$ and the probability for the $n$-th qubit to be in excited
state:

\begin{equation}\label{QbWF1}
    \Psi_Q=\sum\limits_{n,m=1}^N {\left| n \right\rangle }
\lambda_m R_{n,m}e^{ikx_m}
\end{equation}

\begin{equation}\label{prob1}
\left\langle {n|\Psi_Q } \right\rangle  = \sum\limits_{m=1}^N
\lambda_m R_{n,m}e^{ikx_m}
\end{equation}

We assume that all qubits are arranged in the array from left to
right, so that $x_1$ is the position of the qubit at the left end
of the array and $x_N$ is the qibut's position at its right end.
In this case the photon wavefunction (\ref{Ph4}) outside the array
can be written as:
\begin{equation}\label{1qb6}
    {\Psi _N}(x) = \left\{ \begin{array}{l}
t_N{e^{ikx}}\mspace{20mu}{\rm{                 }}(x > x_N)\\
{e^{ikx}} + r_N{e^{ - ikx}}\mspace{20mu}{\rm{     }}(x < x_1)
\end{array} \right.
\end{equation}

where the transmission and reflection coefficients are as follows

\begin{equation}\label{tgen}
   t_N=1-i\hbar\sum\limits_{m,n = 1}^N
     (\Gamma_m\Gamma_n)^{1/2}{e^{ikx_m}R_{m,n}e^{-ikx_n}}
\end{equation}
\begin{equation}\label{rgen}
   r_N=-i\hbar\sum\limits_{m,n = 1}^N
     (\Gamma_m\Gamma_n)^{1/2}{e^{ikx_m}R_{m,n}e^{ikx_n}}
\end{equation}

The conservation of the energy flux requires the additional
condition for $t$ and $r$:
\begin{equation}\label{tr}
    |t_N|^2+|r_N|^2=1
\end{equation}
The expressions (\ref{tgen}) and (\ref{rgen}) are of general
nature and they form the basis for the calculation of microwave
transmission and reflection in particular cases.

\section{Transmission, reflection, and photon mediated interactions in the qubit system}

\subsection{One qubit in a waveguide}\label{SS1}
In this case, in subspace Q there is the only vector $|1\rangle$.
We also assume that the qubit is located at the point $x=0$. From
(\ref{tgen}) and (\ref{rgen}) we obtain:

\[t_1=1-i\hbar \Gamma R_{11}\]
\[r_1=-i\hbar \Gamma R_{11}\]
where
\begin{equation}\label{R11}
 R_{11}=\langle 1|\frac{1}{E-H_{eff}}| 1 \rangle=\frac{1}{E-\langle 1|
H_{eff}|1\rangle}
\end{equation}

The running energy $E$ in (\ref{R11}) is the energy of incident
photon plus the energy of the qubit in the ground state,
$E=\hbar\omega-\hbar\Omega/2$. From (\ref{Hij2}) we also have:
\[ \langle 1|H_{eff}|1\rangle=\frac{\hbar\Omega}{2}-i\hbar\Gamma \]

\begin{figure}
  \includegraphics[height=.2\textheight]{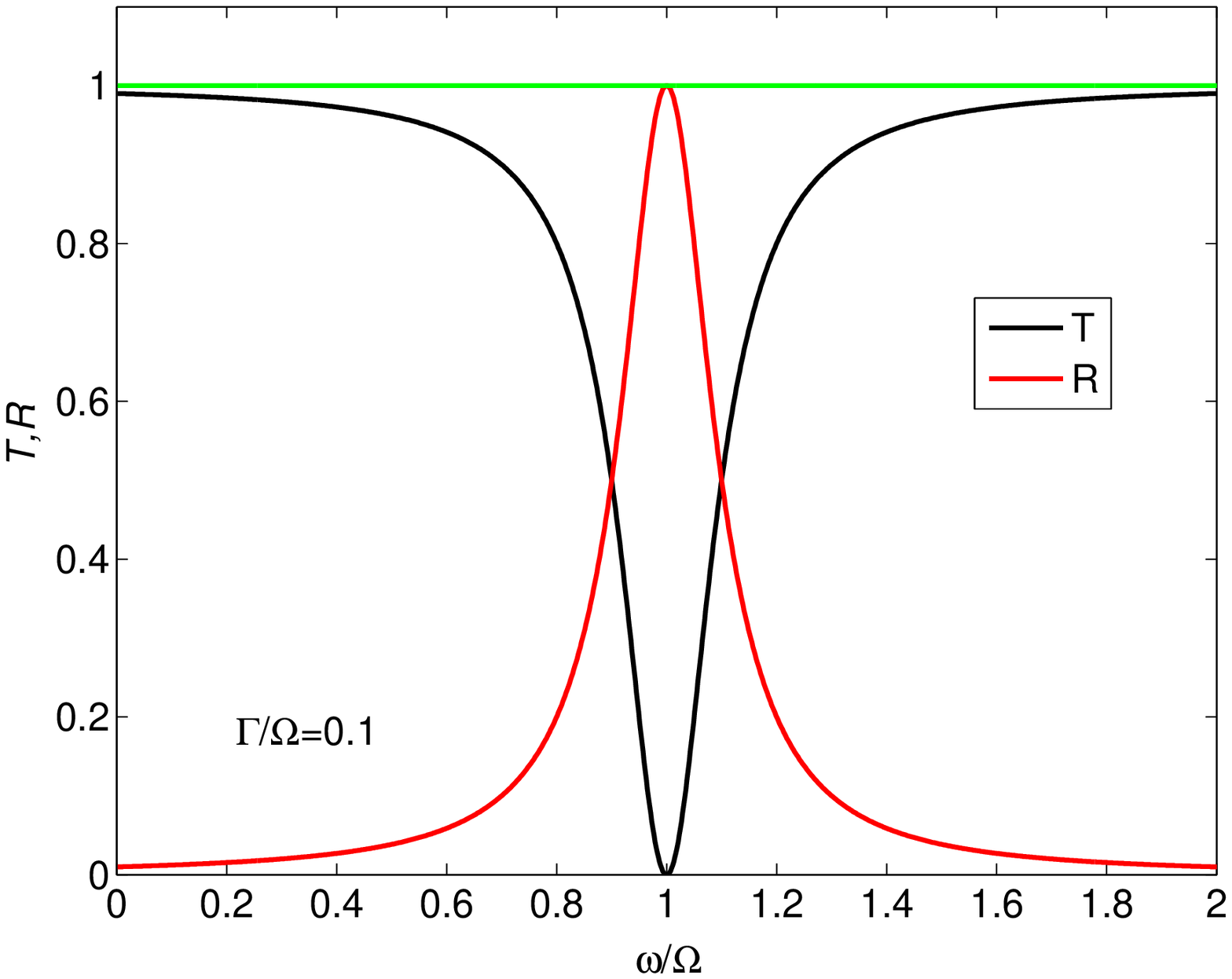}
  \caption{Color online. Transmittance $T=|t|^2$ and reflectance $R=|r|^2$ as functions of
   $\omega/\Omega$ for one qubit in a waveguide. The width of resonance is $2\Gamma$.}\label{fig2}
\end{figure}

Hence, for $t$ and $r$ we finally obtain:
\begin{equation}\label{1qb4}
  t_1=\frac{\omega-\Omega}{\omega-\Omega+i\Gamma}
\end{equation}
\begin{equation}\label{1qb5}
  r_1=\frac{-i\Gamma}{\omega-\Omega+i\Gamma}
\end{equation}
The plots of microwave transmission and reflection are shown in
Fig.\ref{fig2}. At resonance the signal transmission is zero.

The expressions (\ref{1qb4}), (\ref{1qb5}) coincide with those
obtained in \cite{Shen05b} where one qubit problem has been solved
in a configuration space.

From (\ref{1qb4}) and (\ref{1qb5}) we see that $t-r=1$. Unlike the
general condition (\ref{tr}) it is valid only for one qubit and
reflects the continuity of the wavefunction (\ref{1qb6}) at the
point $x=0$.

The probability for the qubit to be excited is given by
(\ref{prob1}) with $n=1$:
\begin{equation}\label{prob1a}
    \langle 1|\Psi_Q\rangle=\lambda R_{11}=\frac{\lambda}{\hbar}\frac{1}{\omega-\Omega+i\Gamma}
\end{equation}

\subsection{Two qubits in a waveguide}\label{B}

\subsubsection{Spectral properties of effective
Hamiltonian}\label{B1}

Here we consider the first nontrivial example that exhibits
superradiant transition: the two noninteracting qubits in a
waveguide . The qubits are positioned at the points $x_1=-d/2$ and
$x_2=+d/2$, respectively, with a distance $d$ between them. The Q-
subspace is formed by two state vectors
$|1\rangle\equiv|e_1,g_2,0\rangle$ and
$|2\rangle\equiv|g_1,e_2,0\rangle$. According to (\ref{Hij2}) the
matrix of effective Hamiltonian is as follows:

\begin{equation}\label{2qb1}
{H_{eff}} = \left( {\begin{array}{*{20}{c}}
{\varepsilon  - i\hbar {\Gamma _1}}&{ - i\hbar \sqrt {{\Gamma _1}{\Gamma _2}} {e^{ikd}}}\\
{ - i\hbar \sqrt {{\Gamma _1}{\Gamma _2}} {e^{ikd}}}&{ -
\varepsilon  - i\hbar {\Gamma _2}}
\end{array}} \right)
\end{equation}

where $\varepsilon=\frac{\hbar}{2}(\Omega_1-\Omega_2)$, and
$\Gamma_i, (i=1,2)$ are defined in (\ref{SE}).

From the matrix (\ref{2qb1}) we can find the complex energies of
the Q- system from the equation (\ref{2qb2}), where $\rm E  =
{\rm{ }}\hbar \widetilde \omega  - \frac{\hbar }{2}\left( {{\Omega
_1} + {\Omega _2}} \right)$. For two qubit case the equation
(\ref{2qb2}) gives two poles in the complex $\widetilde{\omega}$
plane as the function of physical frequency $\omega$.

\begin{eqnarray}\label{2qb3}
\tilde \omega  = \frac{{{\Omega _1} + {\Omega _2}}}{2} - i\frac{{{\Gamma _1} + {\Gamma
_2}}}{2}\nonumber\pm\\[3pt]
 \sqrt {\frac{1}{4}{{\left( {{\Omega _1} - {\Omega _2}
 + i[{\Gamma _2} - {\Gamma _1}]} \right)}^2}- {\Gamma _1}{\Gamma
 _2}{e^{2ikd}}}
\end{eqnarray}
From this expression it follows that the positions of resonances
and their widths depend on the frequency $\omega$ of incident
photon ($k=\omega/v_g$). This is a common feature of non Markovian
behavior if the number of qubits is more than one.

For identical noninteracting qubits $\Omega_1=\Omega_2=\Omega$,
$\Gamma_1=\Gamma_2=\Gamma$ we obtain from (\ref{2qb3})
\begin{equation}\label{2qb4}
    \widetilde \omega  = \Omega  - i\Gamma  \pm i\Gamma {e^{ikd}}
\end{equation}
In the complex $\widetilde{\omega}$  plane the roots are as
follows
\begin{equation}\label{2qb5}
    {\mathop{\rm Re}}\widetilde \omega  \equiv \widetilde E =
\Omega  \mp \Gamma \sin kd
\end{equation}
\begin{equation}\label{2qb6}
    {\mathop{\rm Im}} \widetilde \omega  \equiv \widetilde \Gamma  =
     - \Gamma \left( {1 \mp \cos kd} \right)
\end{equation}
From (\ref{2qb5}) and (\ref{2qb6}) we obtain the relation between
the real and imaginary part of the roots
\begin{equation}\label{2qb7}
    {\left( {\widetilde E - \Omega } \right)^2} + {\left( {\widetilde \Gamma
    + \Gamma } \right)^2} = {\Gamma ^2}
\end{equation}
It is remarkable that this relation does not depend on the $k$, i.
e., on the running frequency $\omega$. In the plane
$(\widetilde\Gamma,\widetilde E)$ all solutions of eq.
(\ref{2qb7}) lie at the circle centered in the point
$-\Gamma,\Omega$ with radius equal to $\Gamma$.
\begin{figure}
 \includegraphics[height=.2\textheight, angle=-90]{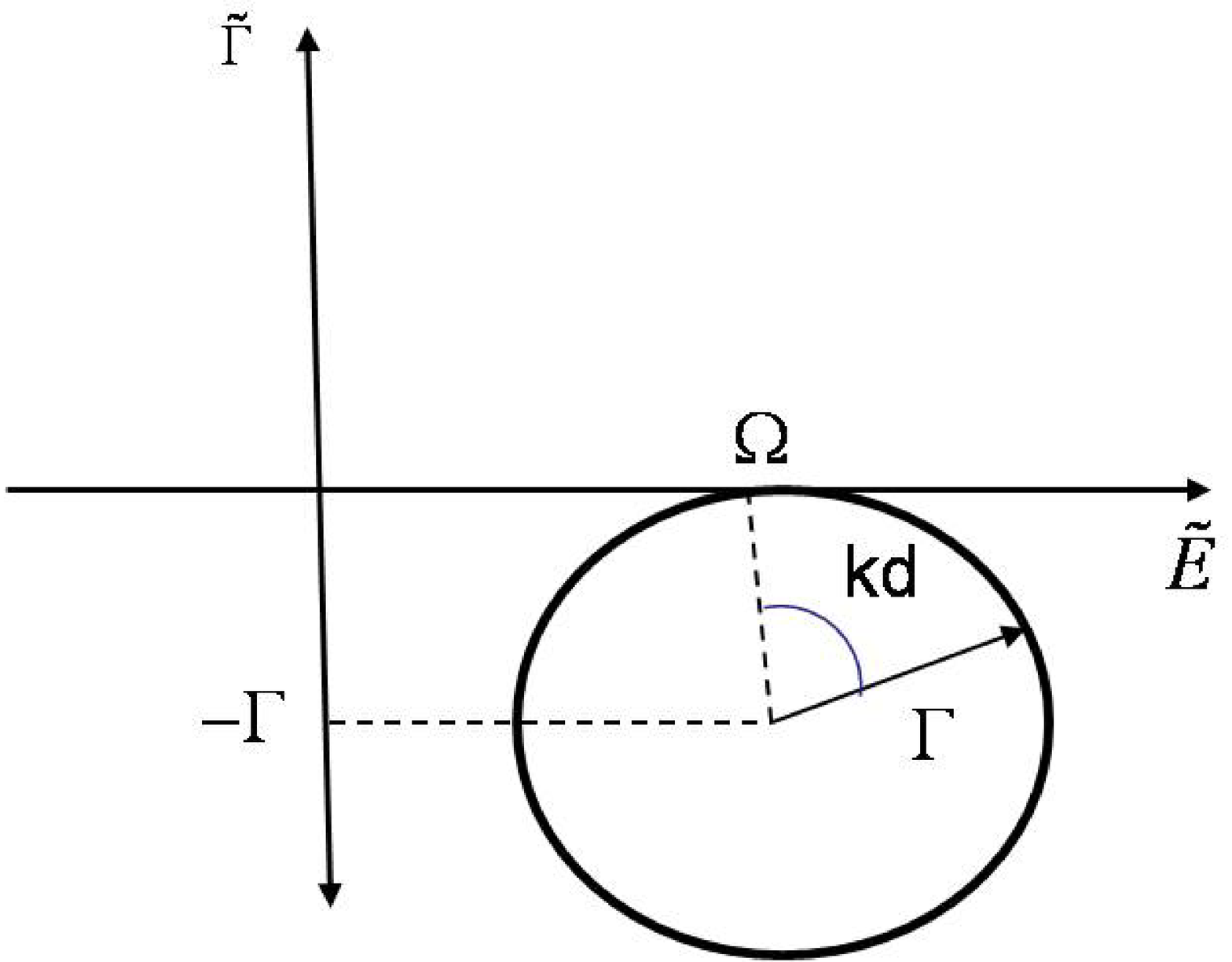}
  \caption{The relation between real and imaginary part of the roots
  in the complex $\widetilde{\omega}$ plane. For every $kd$ there are two roots which lie
  on the circle
at the opposite points} \label{fig1}
\end{figure}
In the long wavelength limit $(kd<<1)$ we obtain from (\ref{2qb4})
two poles
\[{\widetilde \omega _ + } = \Omega  + \Gamma \frac{\omega
}{{{{\rm{v}}_g}}}d - i2\Gamma\]

\[{\widetilde \omega _ - } = \Omega  - \Gamma \frac{\omega
}{{{{\rm{v}}_g}}}\]

We see that in this approximation one of the states absorbs the
width of two qubits. With the increase of $\Gamma$  two states
repel each other. This is also holds if $kd$ is integer multiple
of $\pi$: one state becomes stationary while the width of the
other state is $2\Gamma$. However, this case is valid only for
particular values of the running frequency $\omega_n=\pi nv_g/d$,
$(n=1,2...)$.

\subsubsection{Calculation of microwave transmission}\label{B2}

In order to find transmission and reflection factors $t$ and $r$,
it is necessary to calculate the matrix $R_{m,n}$, ($m,n=1,2$)
which is the inverse of the matrix $(E-H_{eff})_{m,n}$, where the
physical energy $E=\hbar\omega-\hbar(\Omega_1+\Omega_2)/2$, and
the elements of the matrix $(H_{eff})_{m,n}$ are given in
(\ref{2qb1}). Direct calculations yield for $R_{m,n}$ the
following result:
\begin{equation}\label{Rmn1}
    {R_{m,n}} = \frac{1}{{\hbar {D_2}(\omega )}}\left( {\begin{array}{*{20}{c}}
{\omega  - {\Omega _2} + i{\Gamma _2}}&{ - i\sqrt {{\Gamma _1}{\Gamma _2}} {e^{ikd}}}\\
{ - i\sqrt {{\Gamma _1}{\Gamma _2}} {e^{ikd}}}&{\omega  - {\Omega
_1} + i{\Gamma _1}}
\end{array}} \right)
\end{equation}
where
\begin{equation}\label{D2}
    {D_2}(\omega ) = \left[ {\omega  - {\Omega _2} + i{\Gamma _2}} \right]
    \left[ {\omega  - {\Omega _1} + i{\Gamma _1}} \right] + {\Gamma _1}{\Gamma _2}{e^{i2kd}}
\end{equation}

Finally, according to prescriptions in (\ref{tgen}) and
(\ref{rgen}) we obtain  $t$ and $r$ in terms of running frequency
$\omega$:
\begin{equation}\label{2qbt}
t_2   = \frac{{(\omega  - {\Omega _1})(\omega  - {\Omega
_2})}}{{\left[ {\omega  - {\Omega _2} + i{\Gamma _2}}
\right]\left[ {\omega  - {\Omega _1} + i{\Gamma _1}} \right] +
{\Gamma _1}{\Gamma _2}{e^{2ikd}}}}
\end{equation}
\begin{equation}\label{2qbr}
r_2  =  - i\frac{{\left\{ {{e^{ikd}}{\Gamma _1}\left[ {\omega  -
{\Omega _2} - i{\Gamma _2}} \right]} \right. + \left. {{e^{ -
ikd}}{\Gamma _2}\left[ {\omega  - {\Omega _1} + i{\Gamma _1}}
\right]} \right\}}}{{\left[ {\omega  - {\Omega _2} + i{\Gamma _2}}
\right]\left[ {\omega  - {\Omega _1} + i{\Gamma _1}} \right] +
{\Gamma _1}{\Gamma _2}{e^{2ikd}}}}
\end{equation}

For identical qubits we obtain from (\ref{2qbt}), (\ref{2qbr}):

\begin{equation}\label{2qbtid}
t_2 = \frac{{{{(\omega  - \Omega )}^2}}}{D_2^{id}(\omega)}
\end{equation}
\begin{equation}\label{2qbrid}
r_2 =  - i\frac{{2\Gamma [(\omega  - \Omega )\cos kd + \Gamma \sin
kd]}}{D_2^{id}(\omega)}
\end{equation}
where
\begin{equation}\label{2id}
D_2^{id}(\omega) = {{{{(\omega - \Omega  + i\Gamma )}^2} + {\Gamma
^2}{e^{2ikd}}}}
\end{equation}

As is seen from these expressions the form of the transmission and
reflection spectra depend on the inter qubit distance $d$. In the
long wavelength limit we obtain from (\ref{2qbtid}) and
(\ref{2qbrid}):
\begin{equation}\label{tlw}
t_2= \frac{{\omega  - \Omega }}{{\omega  - \Omega  + i2\Gamma }}
\end{equation}
\begin{equation}\label{rlw}
r_2 = \frac{-i2\Gamma}{{\omega  - \Omega  + i2\Gamma }}
\end{equation}
The expressions (\ref{tlw}), (\ref{rlw}) are identical to the one
qubit case (\ref{1qb4}), (\ref{1qb5}) with the only exception. For
two identical qubits the resonance width is twice the resonance
width for one qubit, which is clear signature of superradiance
transition which corresponds to a coherent symmetric superposition
$({\Psi_Q) _S} = a\left( {\left| 1 \right\rangle  + \left| 2
\right\rangle } \right)$, whete the quantity $a$ is given in the
next subsection.

Below we show several plots of transmission and reflection
amplitudes for different values of $k_0d$, where $k_0=\Omega/v_g$.
The plots are calculated for two identical qubits from
(\ref{2qbtid}) and (\ref{2qbrid}). The points of the full
transmission corresponds to zeros of the numerator of the
expression (\ref{2qbrid}). Hence, the full transmission is
observed at the points where the reflection is exactly equal to
zero.

\begin{figure}[h]
\begin{minipage}[h]{0.47\linewidth}
\center{\includegraphics[width=1\linewidth]{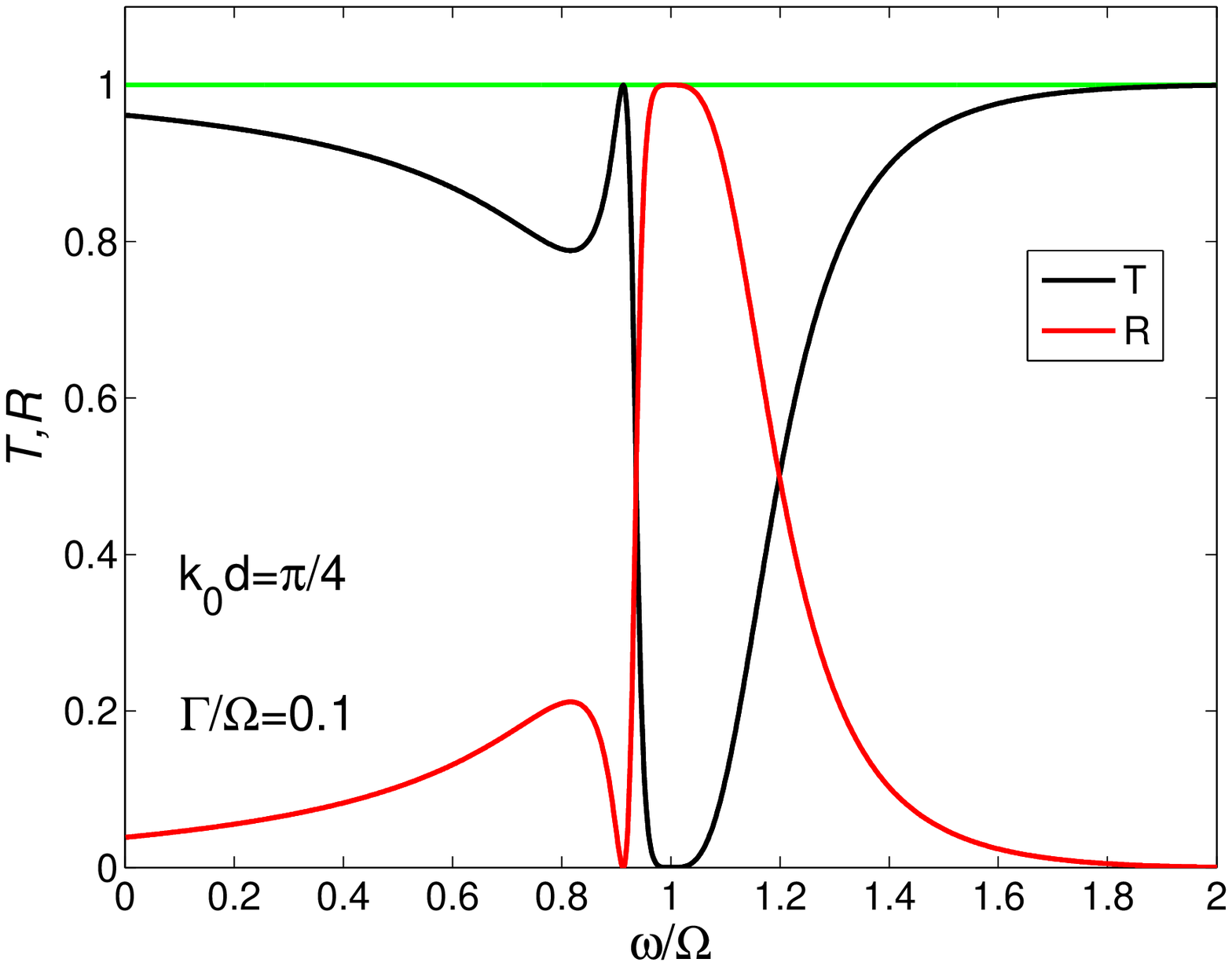}} a) \\
\end{minipage}
\hfill
\begin{minipage}[h]{0.47\linewidth}
\center{\includegraphics[width=1\linewidth]{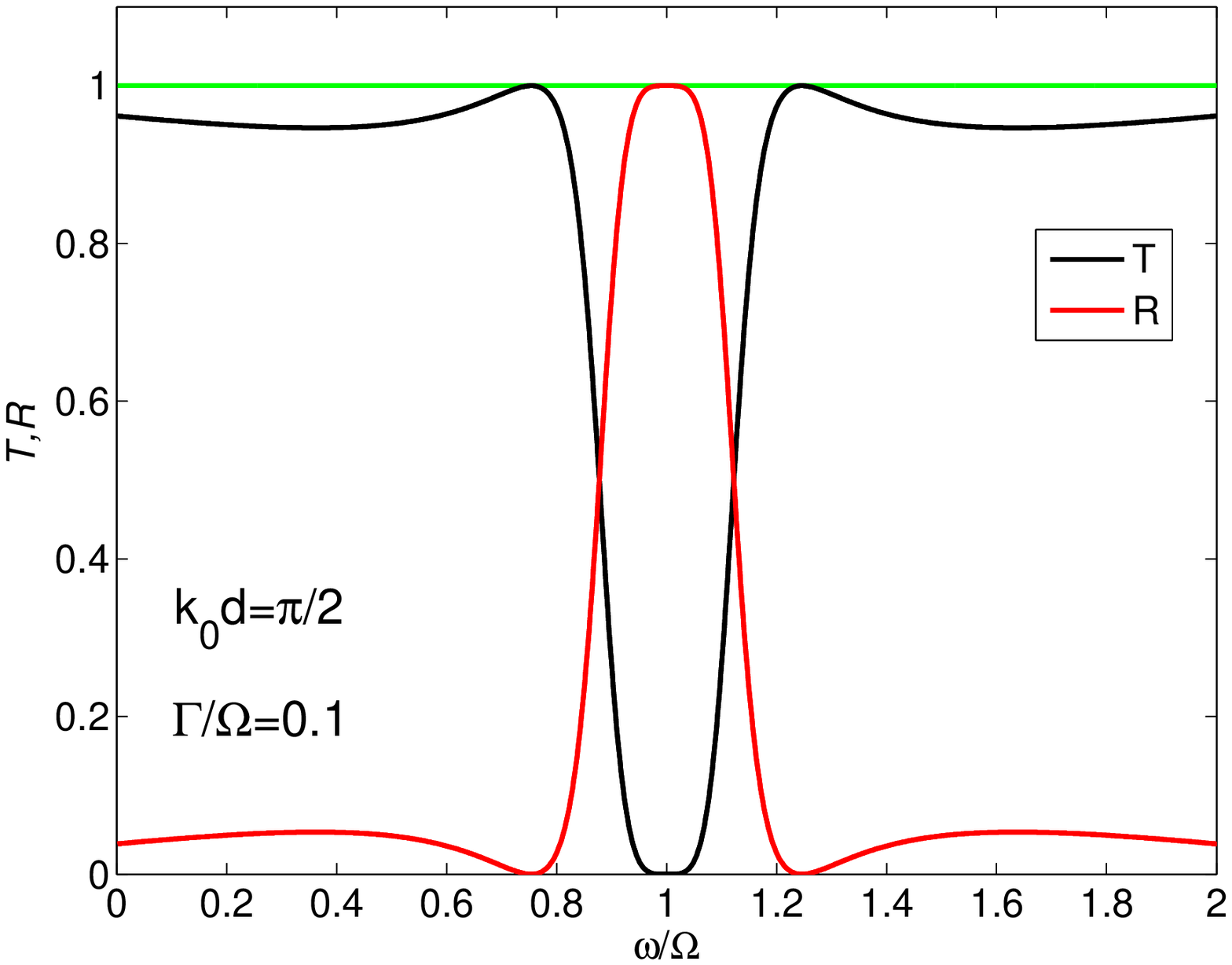}} \\b)
\end{minipage}
\begin{minipage}[h]{0.47\linewidth}
\center{\includegraphics[width=1\linewidth]{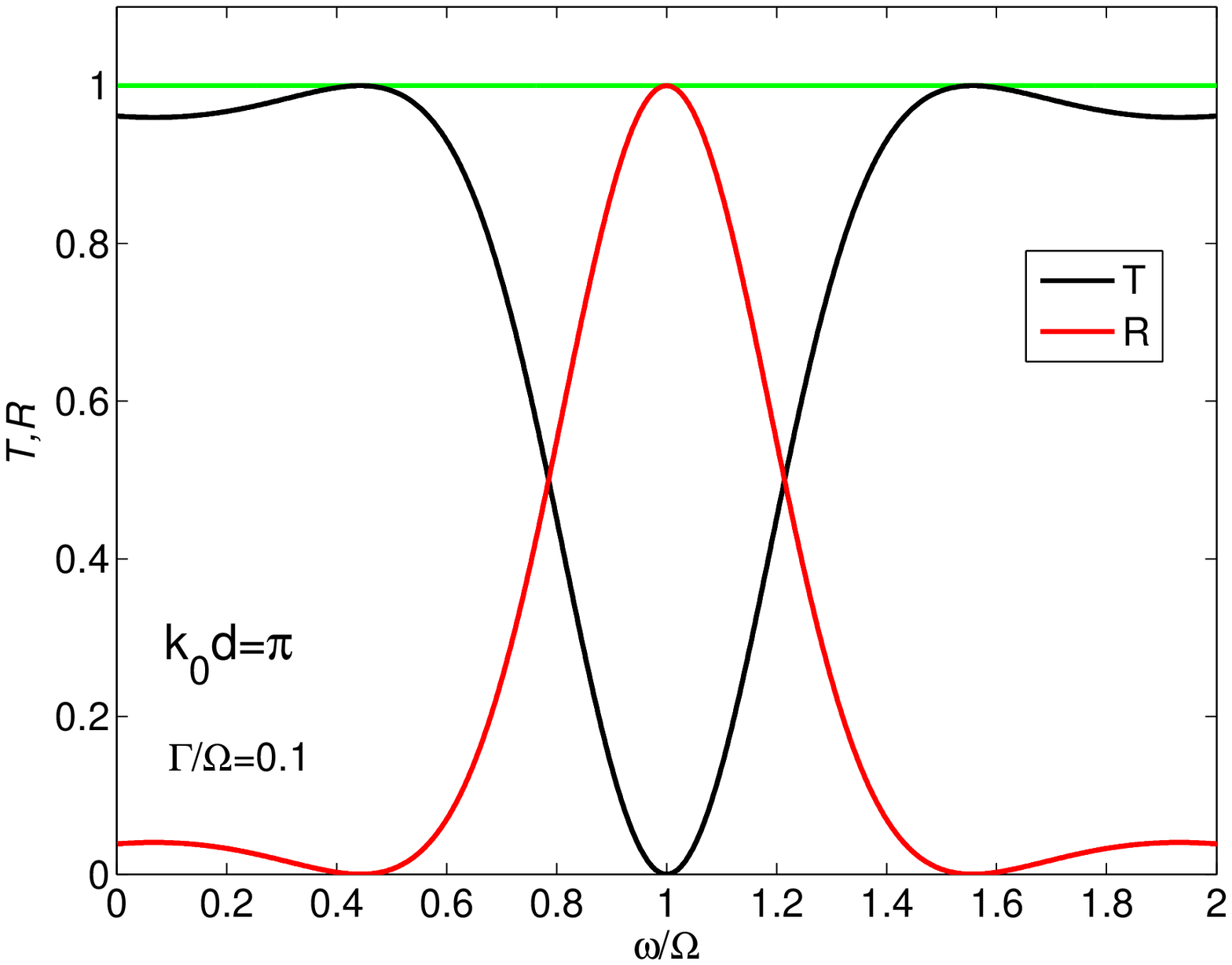}} c) \\
\end{minipage}
\hfill
\begin{minipage}[h]{0.47\linewidth}
\center{\includegraphics[width=1\linewidth]{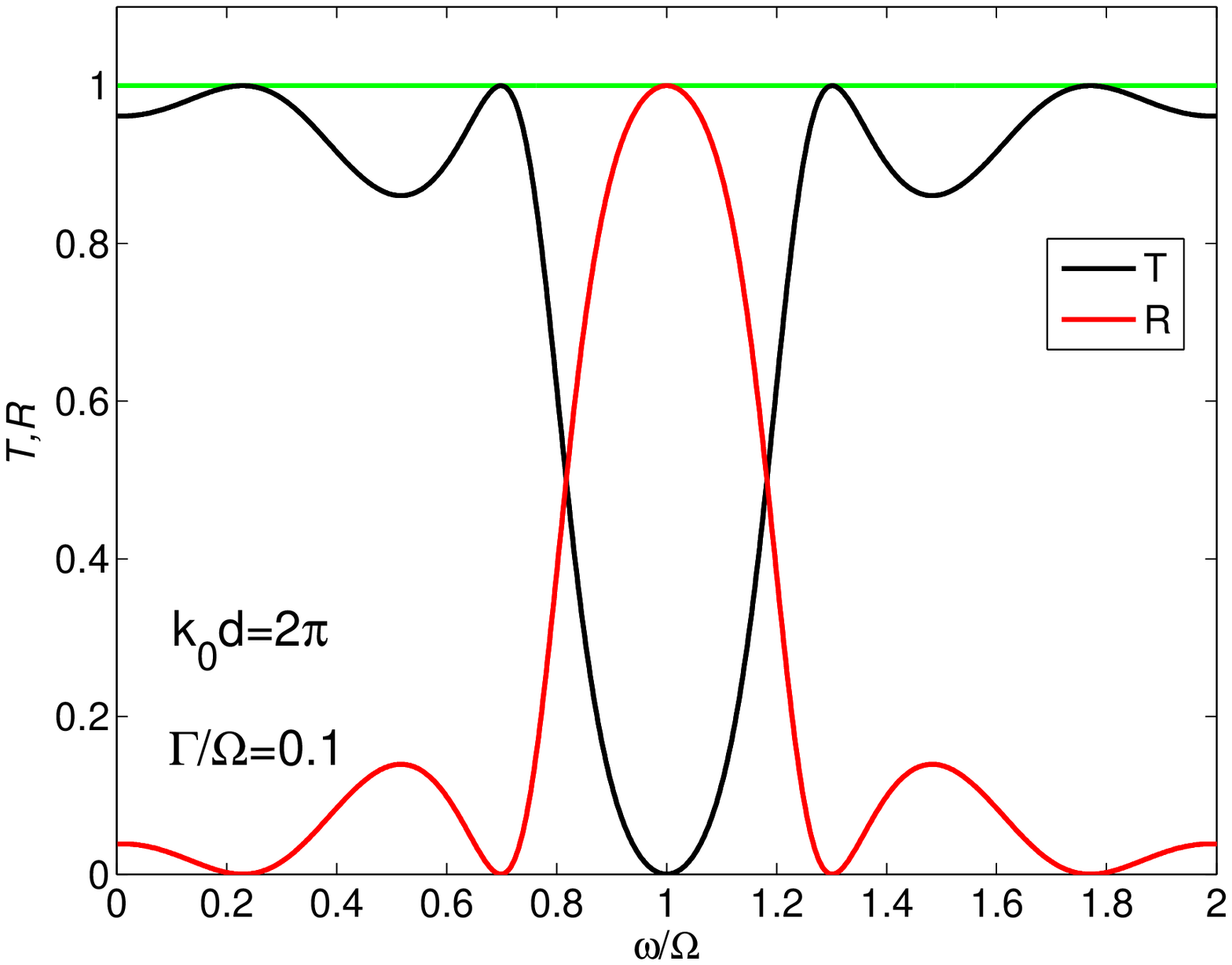}} d) \\
\end{minipage}
\caption{Color online. The dependence of transmission (black) and
reflection (red) amplitudes on the frequency of incident photon,
$\omega/\Omega$ for different values of $k_0d$ for two identical
qubits.} \label{2qbdata}
\end{figure}

\subsubsection{Photon mediated entanglement of two qubits}

For two qubits the structure of the function $\Psi_Q$
(\ref{QbWF1}) within a subspace of qubit states $|1\rangle$ and
$|2\rangle$ is a linear superposition of the two two-qubit states
$\Psi_Q  = a\left| 1 \right\rangle  + b\left| 2 \right\rangle$,
where, in general, $a$ and $b$ depend on the physical frequency
$\omega$. For two identical qubits we obtain a general expression
which describes the frequency dependent entanglement of two
two-qubit states:
\begin{multline}\label{entangl}
    \Psi_Q=\frac{\lambda
    e^{ikd/2}}{D_2^{id}(\omega)}\left([(\omega-\Omega+i\Gamma)e^{-ikd}-i\Gamma
    e^{ikd}]|1\rangle \right.\\ +\left.(\omega-\Omega)|2\rangle\right)
\end{multline}

In the long wavelength limit $kd<<1$ the maximally entangled
superradiant state which corresponds to a coherent symmetric
superposition is formed:

\begin{equation}\label{sym_entangl}
    ({\Psi_Q) _S} = \frac{\lambda}{\omega-\Omega+2i\Gamma}
    \left( {\left| 1 \right\rangle  + \left| 2
\right\rangle } \right)
\end{equation}

The transmission and reflection in this case are given by the
expressions (\ref{tlw}) and (\ref{rlw}). The resonance line of
superradiant state is directly given as the line of reflection
factor (\ref{rlw}).

However, for arbitrary values of $kd$ maximally entangled states
are formed only for particular values of the frequency $\omega$.
For example, if $kd\equiv\omega d/v_g=n\pi$ ($n=1,2,...$) we
obtain from (\ref{entangl}) the expression
\begin{equation}\label{entangl1}
    \Psi_Q=\frac{\lambda
    i^{n}}{\omega_n-\Omega+2i\Gamma}\left[(-1)^n|1\rangle+|2\rangle\right]
\end{equation}
where $\omega_n=n\pi v_g/d$.

For on resonant excitation ($\omega=\Omega$) and $k_0d\neq n\pi$,
where $k_0=\Omega/v_g$ we get from (\ref{entangl}) unentangled
state $\Psi_Q=i\left(\lambda/{\Gamma}\right) e^{-ik_0d/2}
|1\rangle$. In this case we observe a full reflection with only
the first qubit being excited.

\subsubsection{Resonances in two- qubit system}

As it follows from the results of subsection \ref{3_2} the
resonances (their energies and widths) in multi-qubit system are
given by the roots of equation (\ref{2qb2}). The widths of these
resonances define, in general, the decay rates of the qubit
wavefunction $\Psi_Q$ (\ref{QbWF}).

For two qubits these roots, which are labelled below as
$\widetilde{\omega}_1, \widetilde{\omega}_2$, are given in
(\ref{2qb3}). The denominator (\ref{D2}) can then be written as
$D_2(\omega)=\left[\omega-\widetilde{\omega}_1(\omega)\right]$
$\left[\omega-\widetilde{\omega}_2(\omega))\right]$. Hence, the
resonance frequencies of the incident photon are given by the
roots of, in general, nonlinear equations
$\omega=\rm{Re}[\widetilde{\omega}_1(\omega)]$,
$\omega=\rm{Re}[\widetilde{\omega}_2(\omega)]$. These equations
imply that the resonance energies
$\rm{Re}[\widetilde{\omega}_1(\omega)]$,
$\rm{Re}[\widetilde{\omega}_2(\omega)]$ (and their widths
$\rm{Im}[\widetilde{\omega}_1(\omega)]$,
$\rm{Im}[\widetilde{\omega}_2(\omega)]$) depend on the frequency
of incident photon, which comes in (\ref{2qb3}) via the the wave
vector $k=\omega/v_g$. This is a general feature of non Markovian
behavior when the photon mediated interaction between qubits is
not instantaneous and the retardation effects have to be included.
In our method the retardation effects are automatically included
since the quantity $kd$ explicitly enters the expressions for the
transmission and reflection factors. Markovian case corresponds to
long wavelength limit, $kd<<1$ when the propagation time of
photons between the qubits can be neglected and, hence, the qubits
interact instantaneously.

It is important that the resonance widths are directly related to
the lifetime of qubit superposition state (\ref{entangl}) since
they define the poles of denominator $D_2^{id}(\omega)$ in the low
half of the complex energy plane.

Below we discuss the experimental detection of resonance
frequencies and their widths. One of the way is to Fourier
transform the data collected in photon-photon correlation
measurements \cite{Zheng13}. However, the experiments of this type
demand serious attention to optimizing both the measuring system
and experimental conditions. Here we suggest to extract resonance
parameters from directly measured transmission data. As an example
we consider here two identical qubits.

The resonance structure of transmission (\ref{2qbtid}) and
reflection (\ref{2qbrid}) is masked by the frequency dependence of
their numerators. This obstacle can be overcome by appropriate
processing of the output transmission data. In order the resonance
peaks to reveal themselves we propose to divide the transmission
(\ref{2qbtid}) by the factor $[(\omega-\Omega)/\Omega]^2$. Thus,
we analyze the spectral function $S(\omega)$, which contains pure
resonance structure:

\begin{equation}\label{resstr}
    S(\omega)=\frac{\Omega^2}{(\omega  - \Omega  + i\Gamma )^2 +
    \Gamma^2e^{2i\frac{\omega}{\Omega}k_0d}}
\end{equation}

\begin{figure}[h]
  \includegraphics[width=0.8\columnwidth]{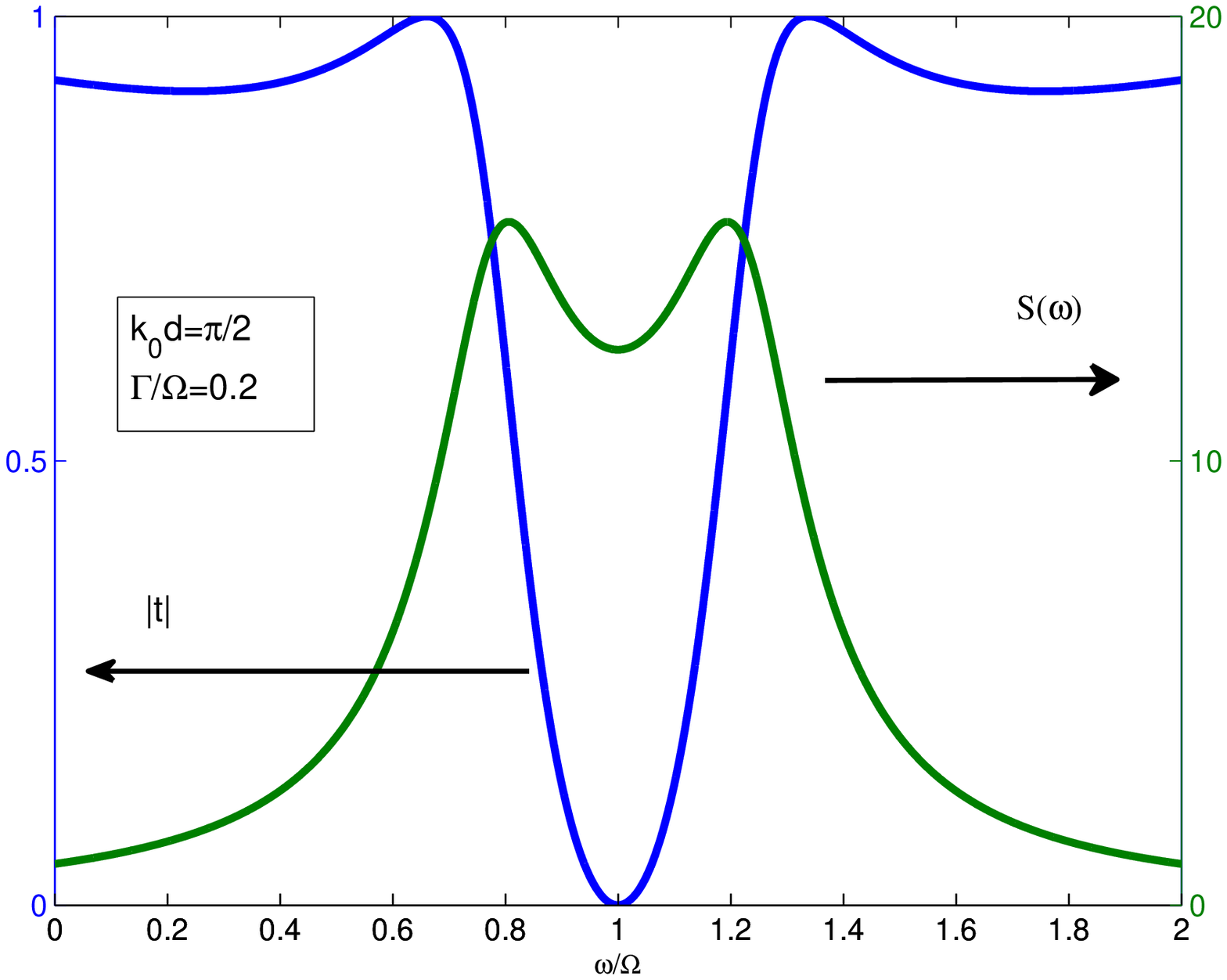}
  \caption{Color online. Frequency dependence of the transmission (left axis, black line)
  and spectral function (right axis, green line)for two identical qubits.
  $k_0d=\pi/2, \Gamma/\Omega=0.2$.}\label{ResPeak}
\end{figure}

The plot of $S(\omega)$ which exhibits two peaks corresponding to
solutions of two equations (see (\ref{2qb5}))
\begin{equation}\label{Pk}
    \omega=\Omega\pm\Gamma\sin\left(\frac{\omega}{\Omega}k_0d\right)
\end{equation}
is shown together with transmission at Fig.\ref{ResPeak}  for
$k_0d=\pi/2, \Gamma/\Omega=0.2$.

\begin{figure}[h]
 \includegraphics[width=0.8\columnwidth]{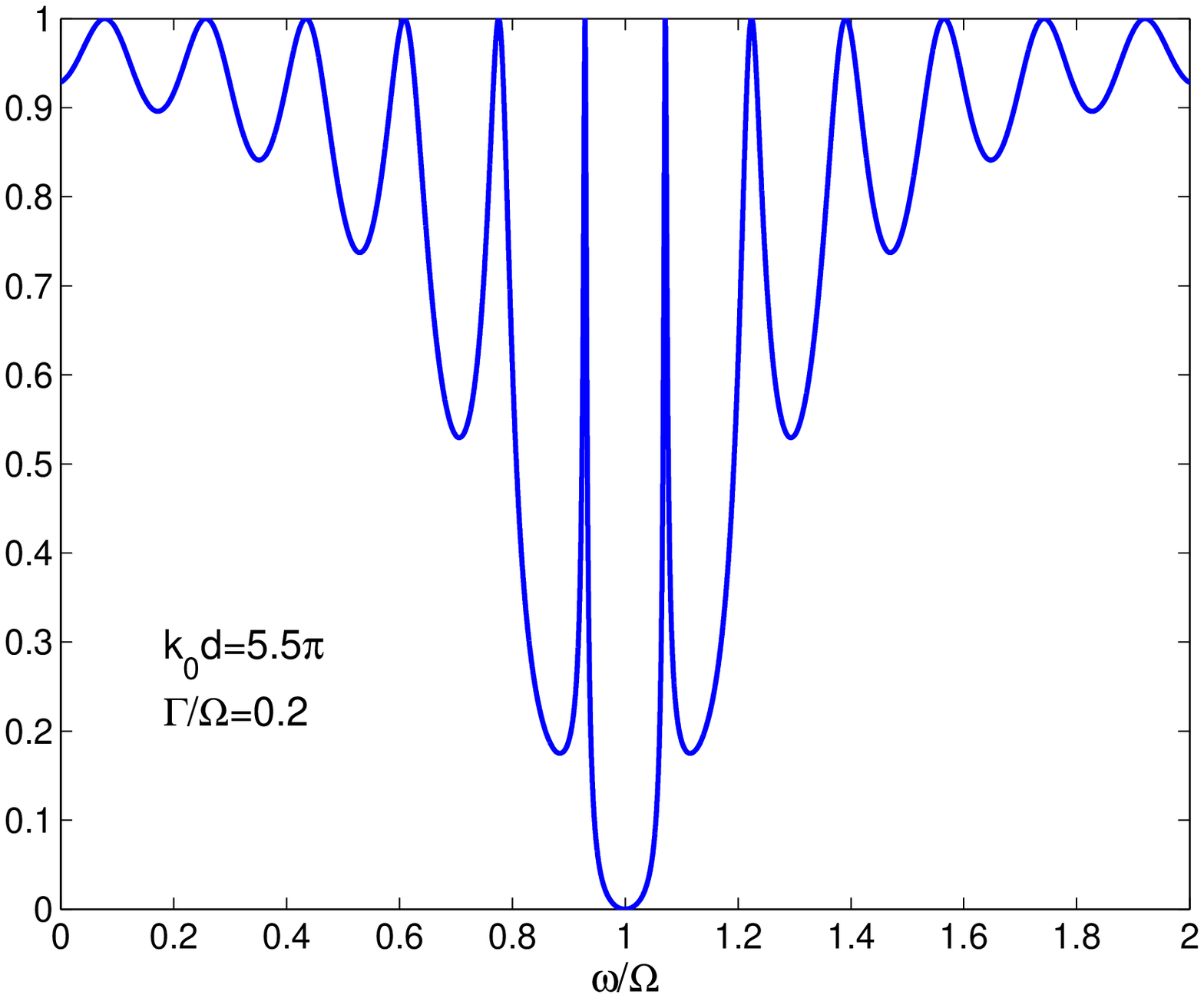}
  \caption{Color online. Transmission pattern for two identical qubits.
  $k_0d=5.5\pi, \Gamma/\Omega=0.2$.}\label{trans_pattern}
\end{figure}

\begin{figure}[h]
  \includegraphics[width=0.8\columnwidth]{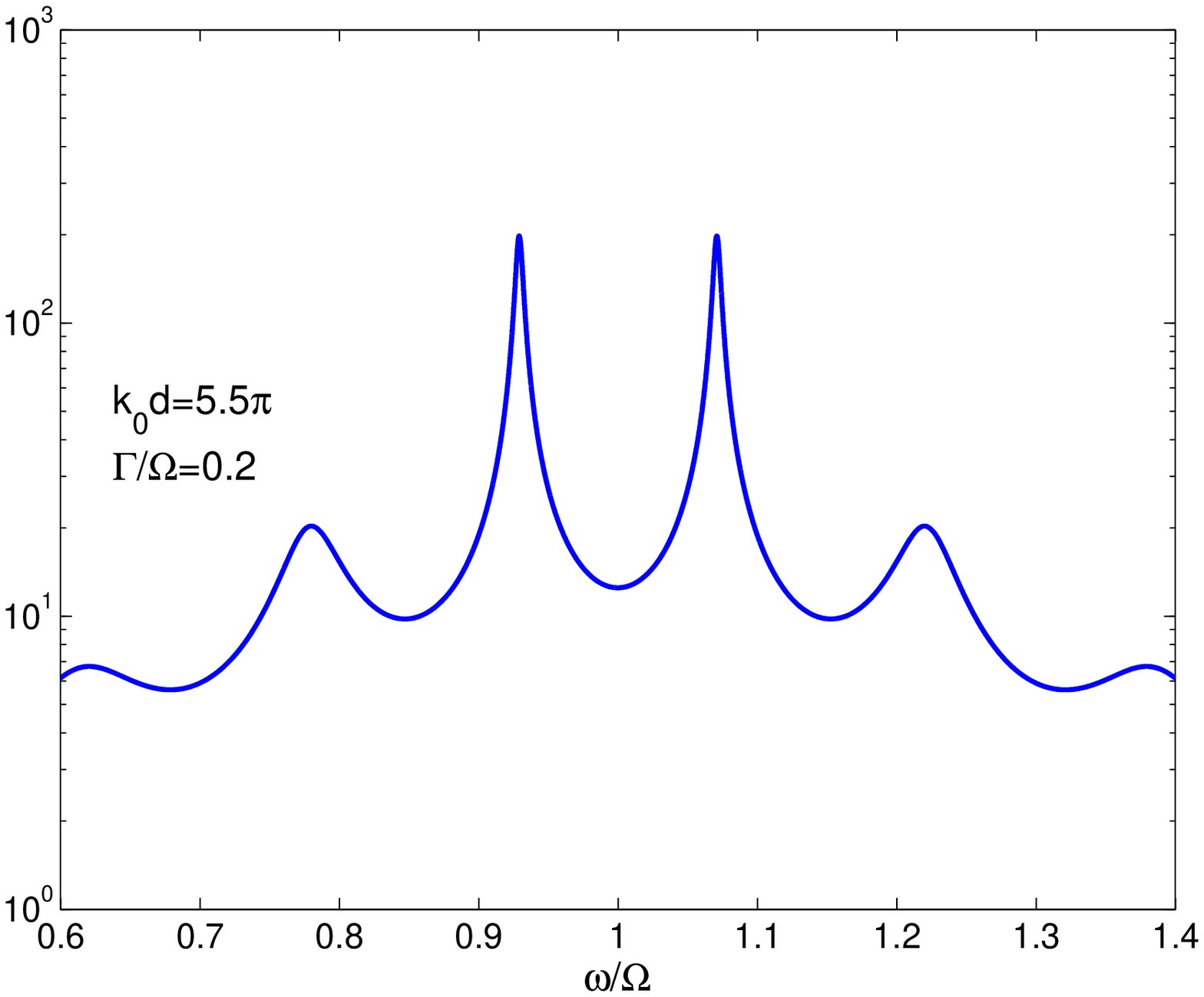}\\
  \caption{Color online. Frequency dependence of the spectral function for two identical qubits.
  $k_0d=5.5\pi, \Gamma/\Omega=0.2$. The $y$-axis is in log scale}\label{ResPeak1}
\end{figure}

We notice that the positions of these peaks at the frequency axis
do not coincide with the points of the full transmission. The
latter points are located exactly where the reflection is zero.
This is well illustrated in Fig.\ref{trans_pattern} and
Fig.\ref{ResPeak1} where the transmission pattern and
corresponding resonance spectrum are shown for $k_0d=5.5\pi,
\Gamma/\Omega=0.2$.

The transmission pattern exhibits $12$ points of the full
transmission within the range of the frequency axis while there
are only six resonances at the frequencies which are given by the
roots of equation (\ref{Pk}): $\omega/\Omega=0.805, 0.866,
0.929,1.070, 1,133, 1.194$ with the corresponding widths
$\widetilde{\Gamma}/\Omega=-0.155, -0.349, -0.013,-0.013, -0.349,
-0.155$ Only four resonances which are sufficiently close to real
axis are visible in Fig. \ref{ResPeak1}. It is worth noting that
there are two resonances with the widths being much smaller the
width of individual qubit. Hence, for $k_0d=5.5\pi$ these two
resonances give the smallest decay rates for two-qubit
superposition state (\ref{entangl}).

We may conclude that as the distance between qubits, $d$ is
increased at fixed $\Gamma$, a number of the roots of (\ref{Pk}),
which for small $\Gamma$'s lie in the narrow range
$\Omega\pm\Gamma$, is also increased, being approximately equal to
$k_0d/\pi$, while the widths of corresponding peaks are decreased.
The latter effect is a direct manifestation of non Markovian
behavior.

\subsubsection{Photon wave function for two qubits in a waveguide}

Photon wave function for two  qubits is calculated from
(\ref{Ph4}) with $x_1=-d/2, x_2=+d/2$ and the matrix $R_{m,n}$
from (\ref{Rmn1}). Outside the qubit array $x>d/2$, $x<-d/2$ the
wavefunction is given by (\ref{1qb6}) with $t$ and $r$ from
(\ref{2qbt}) and (\ref{2qbr}). Below we write the photon
wavefunction for two qubits in the intermediate region
$(-d/2<x<+d/2)$:

\begin{equation}\label{Pw2}
{\Psi _2}(x) = \frac{{(\omega  - {\Omega _1})}}{{{D_2}(\omega
)}}\left( {{e^{ikx}}(\omega  - {\Omega _2} + i{\Gamma _2}) -
i{\Gamma _2}{e^{ikd}}{e^{ - ikx}}} \right)
\end{equation}

It can easily be verified that the wavefunctions (\ref{1qb6}) and
(\ref{Pw2}) are continuous at the points $x=\pm d/2$. At resonance
with the first qubit $(\omega=\Omega_1)$ photon is reflected from
the first qubit and does not penetrate in the inter qubit region
$x>-d/2$. However, at resonance with the second qubit
$(\omega=\Omega_2)$ the wave function $\Psi_2(x)\neq 0$ at inter
qubit region $-d/2<x<d/2$, but $\Psi_2(d/2)=0$ as it follows from
continuity condition.

From (\ref{prob1}) we calculate the probability amplitude for the
first or second qubit to be excited.
\begin{equation}\label{Psi2}
\left\langle {1|{\Psi _Q}} \right\rangle  = \frac{{{e^{ -
ikd/2}}}}{{\hbar {D_2}(\omega )}}\left[ {{\lambda _1}\left(
{\omega  - {\Omega _2} + i{\Gamma _2}} \right) - i{\lambda
_2}\sqrt {{\Gamma _1}{\Gamma _2}} {e^{2ikd}}} \right]
\end{equation}
\begin{equation}\label{Psi1}
\left\langle {2|{\Psi _Q}} \right\rangle  =
\frac{{{e^{ikd/2}}}}{{\hbar {D_2}(\omega )}}\left[ {{\lambda
_2}\left( {\omega  - {\Omega _1} + i{\Gamma _1}} \right) -
i{\lambda _1}\sqrt {{\Gamma _1}{\Gamma _2}} } \right]
\end{equation}
From the definition of $\Gamma$ (\ref{SE}) we may rewrite
(\ref{Psi1}) as:

\[\left\langle {2|{\Psi _Q}} \right\rangle  =
\frac{{{e^{ikd/2}}}}{{\hbar {D_2}(\omega )}}\left[ {{\lambda
_2}\left( {\omega  - {\Omega _1}} \right)} \right]\]

Hence, if the photon is in resonance with the first qubit, the
second qubit remains unexcited. If the photon is in resonance with
the second qubit, the first qubit is unexcited only if $\Omega_2
d/v_g=\pi$.

\subsection{Three qubits in a waveguide}
\subsubsection{Spectral properties of effective Hamiltonian}

Here we consider three noninteracting qubits in a waveguide . The
qubits are positioned at the points $x_1=-d$, $x_2=+d$ and
$x_3=0$, respectively, with a distance $d$ between adjacent
qubits. The Q- subspace is formed by three state vectors
$|1\rangle\equiv|e_1,g_2,g_3,0\rangle$,
$|2\rangle\equiv|g_1,e_2,g_3,0\rangle$ and
$|3\rangle\equiv|g_1,g_2,e_3,0\rangle$. The states $|1\rangle$ and
$|2\rangle$ correspond to qubits located at the points $x=\pm d$,
respectively. The state $|3\rangle$  is for the qubit placed at
the point $x=0$. The $P$-subspace is formed by the vectors
$|k\rangle\equiv|g_1,g_2,g_3,k\rangle$. According to (\ref{Hij2})
the matrix of effective Hamiltonian is as follows:

\begin{equation}\label{Hmtr}
{H_{eff}} = \left( {\begin{array}{*{20}{c}}
{{\varepsilon _1} - i\hbar {\Gamma _1}}&{ - i\hbar \sqrt {{\Gamma _1}{\Gamma _2}} {e^{2ikd}}}&{ - i\hbar \sqrt {{\Gamma _1}{\Gamma _3}} {e^{ikd}}}\\
{ - i\hbar \sqrt {{\Gamma _1}{\Gamma _2}} {e^{2ikd}}}&{{\varepsilon _2} - i\hbar {\Gamma _2}}&{ - i\hbar \sqrt {{\Gamma _2}{\Gamma _3}} {e^{ikd}}}\\
{ - i\hbar \sqrt {{\Gamma _1}{\Gamma _3}} {e^{ikd}}}&{ - i\hbar
\sqrt {{\Gamma _2}{\Gamma _3}} {e^{ikd}}}&{{\varepsilon _3} -
i\hbar {\Gamma _3}}
\end{array}} \right)
\end{equation}
where $\varepsilon_i$ and $\Gamma_i, (i=1,2,3)$ are defined in
(\ref{Hij1}) and (\ref{SE}), respectively.

The roots of this Hamiltonian in the complex frequency plane are
defined by the equation
\[\rm{det}\left(\widetilde{\omega}-\frac{1}{2}\left(\Omega_1+\Omega_2+\Omega_3\right)-H_{eff}/\hbar\right)=0\]
that can be expressed as:

\begin{equation}\label{roots}
\begin{array}{l}
\left( {\tilde \omega  - {\Omega _1} + i{\Gamma _1}} \right)\left( {\tilde \omega  - {\Omega _2} + i{\Gamma _2}} \right)\left( {\tilde \omega  - {\Omega _3} + i{\Gamma _3}} \right)\\
 + \left( {\tilde \omega  - {\Omega _1} + i{\Gamma _1}} \right){\Gamma _2}{\Gamma _3}{e^{2ikd}} + \left( {\tilde \omega  - {\Omega _2} + i{\Gamma _2}} \right){\Gamma _1}{\Gamma _3}{e^{2ikd}}\\
 + \left( {\tilde \omega  - {\Omega _3} - i{\Gamma _3}} \right){\Gamma _1}{\Gamma _2}{e^{4ikd}} = 0
\end{array}
\end{equation}
We note that in general the energies and the widths of resonances
depend on the physical frequency $\omega$ ($k=\omega/v_g$).

For identical qubits ($\Omega_1=\Omega_2=\Omega_3\equiv\Omega$,
$\Gamma_1=\Gamma_2=\Gamma_3\equiv\Gamma$) we obtain from
(\ref{roots})
\begin{equation}\label{roots1}
\begin{array}{l}
{\left( {\tilde \omega  - \Omega  + i\Gamma } \right)^3}
 + 2\left( {\tilde \omega  - \Omega  + i\Gamma } \right){\Gamma ^2}{e^{2ikd}}\\
 + \left( {\tilde \omega  - \Omega  - i\Gamma } \right){\Gamma ^2}{e^{4ikd}} = 0
\end{array}
\end{equation}
In the long wavelength limit we find from (\ref{roots1})

\[{\left( {\tilde \omega - \Omega } \right)^3} + 3i \Gamma
{\left( {\tilde \omega  - \Omega } \right)^2} = 0\]

which gives two resonances with null width and one resonance which
absorbs the widths of all three qubits:

\[\widetilde{\omega}_{1,2}=\Omega;\quad
\widetilde{\omega}_{3}=\Omega-3i\Gamma\]

We obtain the same result if the running frequency $\omega$ in
(\ref{roots1}) corresponds to $kd=n\pi, (n=1,2....)$.

\begin{figure}[h]
 \includegraphics[height=.3\textheight]{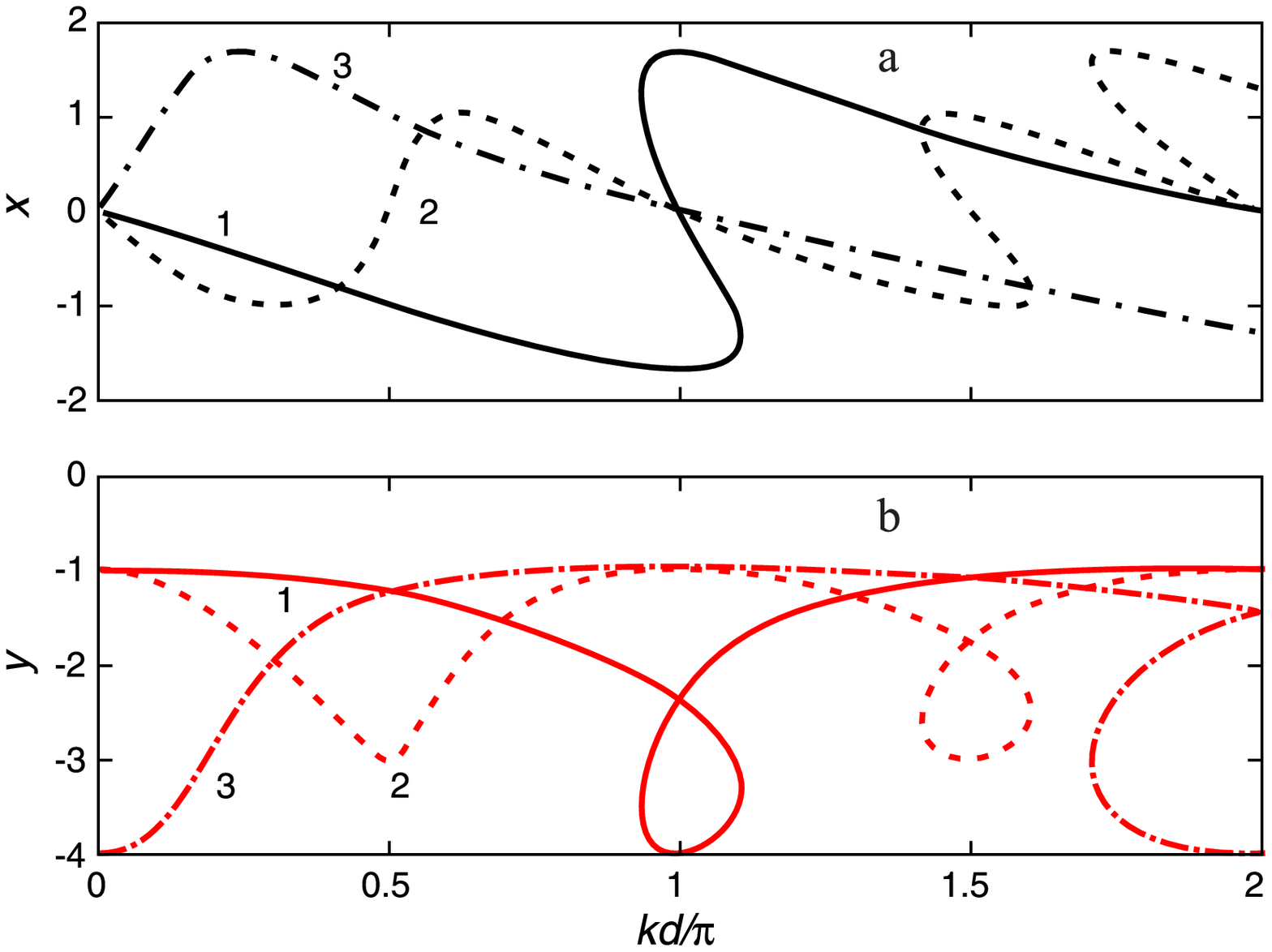}\\
  \caption{Color online. The $kd$- dependence of real (a) and imaginary (b) part of three complex
roots of equation (\ref{roots1}),
$\widetilde{\omega}=\rm{Re}\widetilde{\omega}-i\widetilde{\Gamma}$,
where $x=(\rm{Re}\widetilde{\omega}-\Omega)/\Gamma$,
$y=(\widetilde{\Gamma}-\Gamma)/\Gamma$. The line numbers
correspond to real and imaginary parts of three roots of
Eq.\ref{roots1}.} \label{fig7}
\end{figure}

\begin{figure}[h]
 \includegraphics[height=.3\textheight]{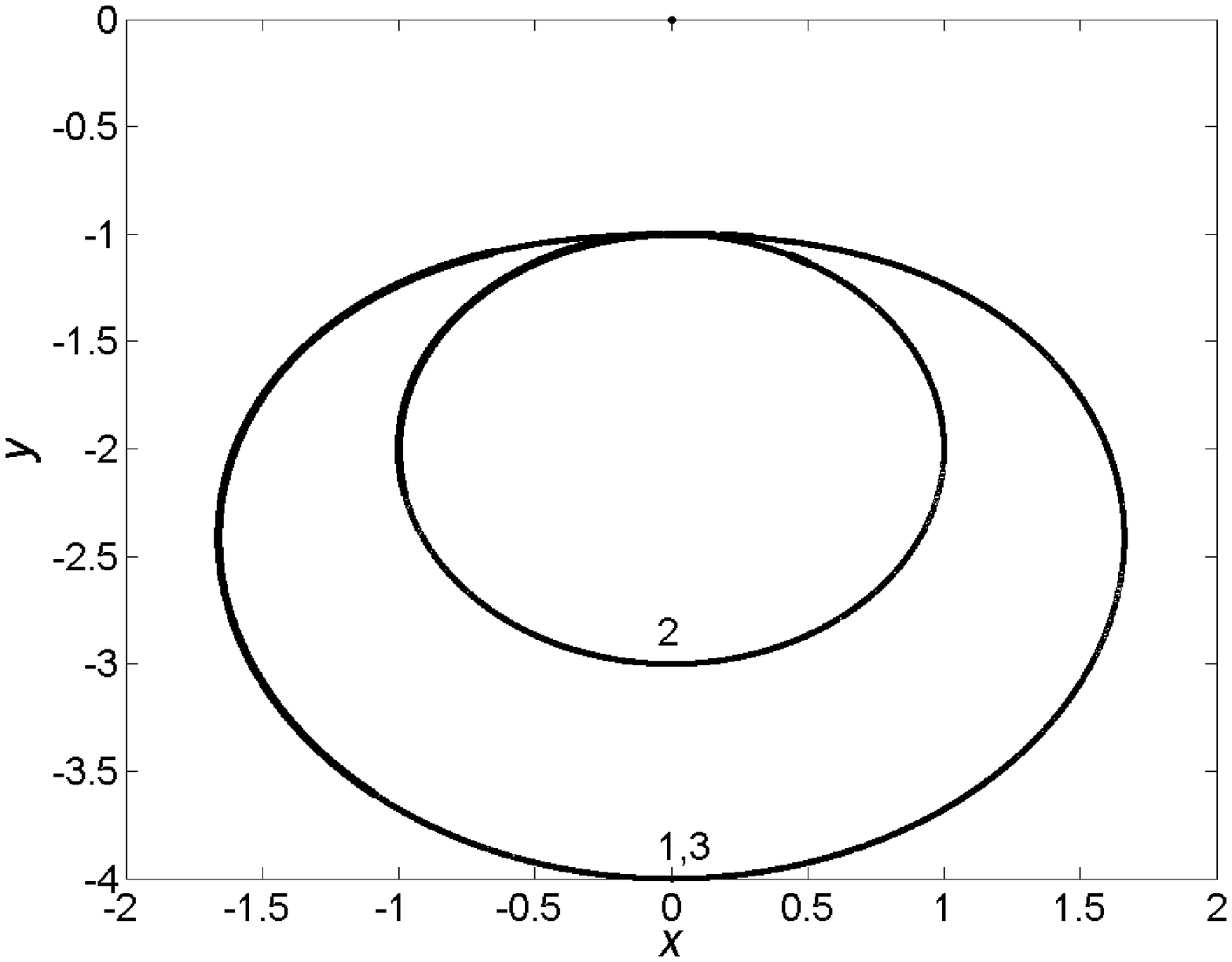}\\
  \caption{The projection of the three roots of Eq. \ref{roots1} to $x,y$
plane, where $x=(\rm{Re}\widetilde{\omega}-\Omega)/\Gamma$,
$y=(\widetilde{\Gamma}-\Gamma)/\Gamma$. Every point on this graph
are merged from black and red points of Fig. \ref{fig7}, which
belong to the same root.}\label{fig8}
\end{figure}

The $kd$- dependence of real and imaginary part of three complex
roots of equation (\ref{roots1}),
$\widetilde{\omega}=\rm{Re}\widetilde{\omega}-i\widetilde{\Gamma}$,
is shown in Fig.\ref{fig7}, where
$x=(\rm{Re}\widetilde{\omega}-\Omega)/\Gamma$,
$y=(\widetilde{\Gamma}-\Gamma)/\Gamma$. In 3D space ($x,y, z=kd$)
three lines of the roots are wound with a variable step onto a
cylindrical surface which has two radii, $\Gamma$ and $3\Gamma/2$.
So that, in the projection to $x,y$ plane these roots form two
circles as shown in Fig.\ref{fig8}. Every point on these circles
is merged from black and red points of Fig. \ref{fig7}, which
belong to the same root.


\subsubsection{Transmission and reflection spectra for three qubits in a waveguide}

Transmission and reflection factors are calculated from
(\ref{tgen}) and (\ref{rgen}), where $m,n=1,2,3$ and $R_{mn}$ is
given in Appendix. In the frequency picture with
  $E=\hbar\omega-\hbar(\Omega_1+\Omega_2+\Omega_3)/2$ we obtain for
  three qubits $t$ and $r$ the following expressions:
\begin{equation}\label{3t}
t_3= \frac{{(\omega  - {\Omega _1})(\omega  - {\Omega _2})(\omega
- {\Omega _3})}}{{D_3\left( \omega  \right)}}
\end{equation}
\begin{equation}\label{3r}
r_3=  - i\frac{{G(\omega )}}{{D_3(\omega )}}
\end{equation}
where
\begin{equation}\label{Ft}
\begin{array}{l}
D_3(\omega ) = \left( {\omega  - {\Omega _1} + i{\Gamma _1}}
\right)\left( {\omega  - {\Omega _2} + i{\Gamma _2}} \right)\left(
{\omega  - {\Omega _3} + i{\Gamma _3}} \right)\\[3pt]
 + \left( {\omega  - {\Omega _1} + i{\Gamma _1}} \right){\Gamma _2}{\Gamma _3}{e^{2ikd}} + \left( {\omega  - {\Omega _2} + i{\Gamma _2}} \right){\Gamma _1}{\Gamma
 _3}{e^{2ikd}}\\[3pt]
 + \left( {\omega  - {\Omega _3} - i{\Gamma _3}} \right){\Gamma _1}{\Gamma _2}{e^{4ikd}}
\end{array}
\end{equation}

\begin{equation}\label{Gr}
\begin{array}{l}
G(\omega ) = {\Gamma _1}(\omega  - {\Omega _2} + i{\Gamma _2})(\omega  - {\Omega _3} + i{\Gamma _3}){e^{ -
2ikd}}\\[3pt]
 + {\Gamma _2}(\omega  - {\Omega _1} + i{\Gamma _1})(\omega  - {\Omega _3} + i{\Gamma
 _3}){e^{2ikd}}\\[3pt]
 + {\Gamma _3}(\omega  - {\Omega _1})(\omega  - {\Omega _2}) + i{\Gamma _2}{\Gamma _3}(\omega  - {\Omega
 _1})\\[3pt]
 + i{\Gamma _1}{\Gamma _3}(\omega  - {\Omega _2}) - 2i{\Gamma _1}{\Gamma _3}(\omega  - {\Omega _2} + i{\Gamma
 _2})\\[3pt]
 - 2i{\Gamma _1}{\Gamma _2}(\omega  - {\Omega _3}){e^{2ikd}} - 2i{\Gamma _3}{\Gamma _2}(\omega  - {\Omega _1}){e^{2ikd}}
\end{array}
\end{equation}

For identical qubits we obtain
\begin{equation}\label{tid}
    t_3 = \frac{{{{(\omega  - \Omega )}^3}}}{{{D_3^{id}}\left( \omega  \right)}}
\end{equation}

\begin{equation}\label{rid}
    r_3 =  - i\frac{{{G_{id}}(\omega )}}{{{D_3^{id}}(\omega )}}
\end{equation}
where $D_3^{id}(\omega)$ is calculated in the section 2 of
Appendix.
\begin{equation}\label{Fid}
\begin{array}{l}
D_3^{id}(\omega) = {\left( {\omega  - \Omega  + i\Gamma }
\right)^3}
 + 2\left( {\omega  - \Omega  + i\Gamma } \right){\Gamma
 ^2}{e^{2ikd}}\\[3pt]
 + \left( {\omega  - \Omega  - i\Gamma } \right){\Gamma ^2}{e^{4ikd}}
\end{array}
\end{equation}
\begin{equation}\label{Gid}
\begin{array}{l}
G_{id}(\omega ) = 2\Gamma [(\omega  - \Omega)^2 -\Gamma ^2]\cos
2kd
+ \Gamma {(\omega  - \Omega )^2} + 2{\Gamma ^3}\\[3pt]
 +4{\Gamma ^2}(\omega  - \Omega )\sin2kd
\end{array}
\end{equation}
In the long wavelength limit $(kd<<1)$ we obtain from  (\ref{tid})
and (\ref{rid})
\begin{equation}\label{tidlw}
t_3 = \frac{{\omega  - \Omega }}{{\omega  - \Omega + 3i\Gamma}}
\end{equation}

\begin{equation}\label{ridlw}
r_3 =  - \frac{{3i \Gamma }}{{\omega  - \Omega
 + 3i\Gamma }}
\end{equation}

We note that the expressions (\ref{tidlw}), (\ref{ridlw}) are
valid for a broad range of frequencies satisfying the condition
$kd<<1$. However, as $k= \omega/v_g$ these expressions are also
valid for $kd=n\pi, n=1,2,..$ but only at the fixed frequencies
$\omega_n=n\pi v_g/d$.

As in the two qubits case (\ref{tlw}), (\ref{rlw}) the expressions
(\ref{tidlw}), (\ref{ridlw}) are similar to the ones for one qubit
(\ref{1qb4}), (\ref{1qb5}) but with the width that is three times
greater.

Below we show several plots of transmission and reflection
amplitudes for different values of $k_0d$, where $k_0=\Omega/v_g$.
The plots are calculated for three identical qubits from
(\ref{tid}) and (\ref{rid}).
\begin{figure}[h]
\begin{minipage}[h]{0.47\linewidth}
\center{\includegraphics[width=1\linewidth]{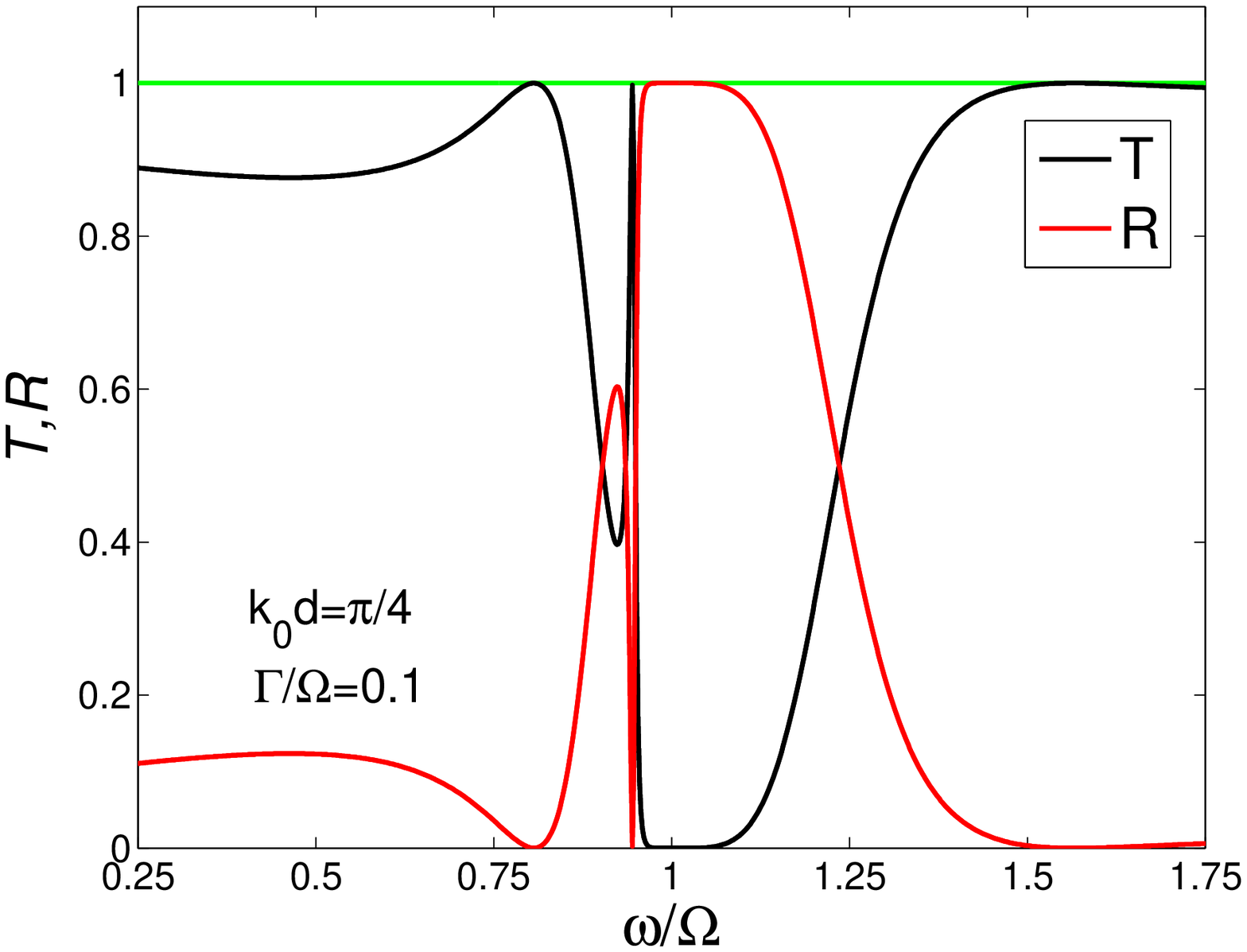}} a) \\
\end{minipage}
\hfill
\begin{minipage}[h]{0.47\linewidth}
\center{\includegraphics[width=1\linewidth]{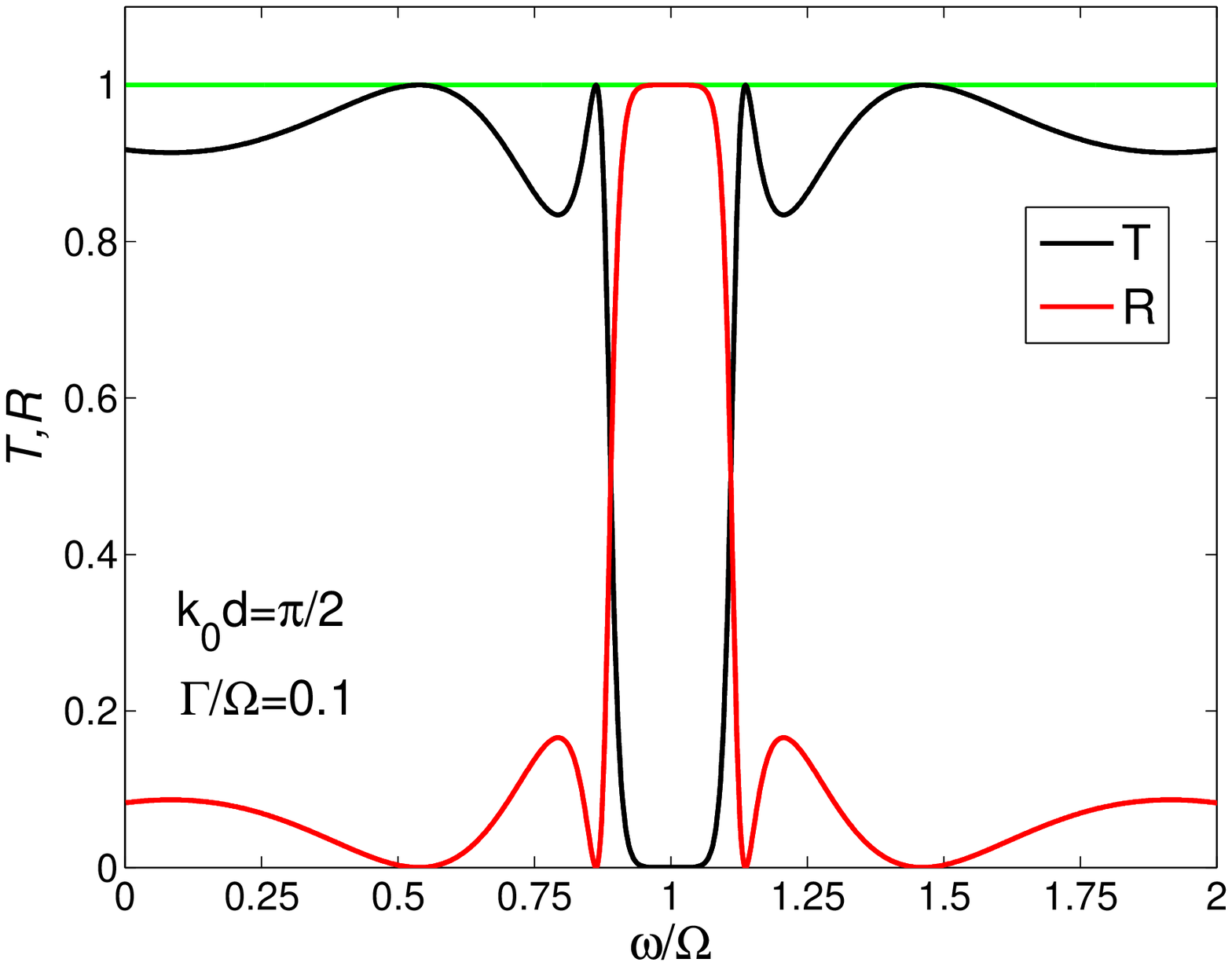}}
\\b)
\end{minipage}
\begin{minipage}[h]{0.47\linewidth}
\center{\includegraphics[width=1\linewidth]{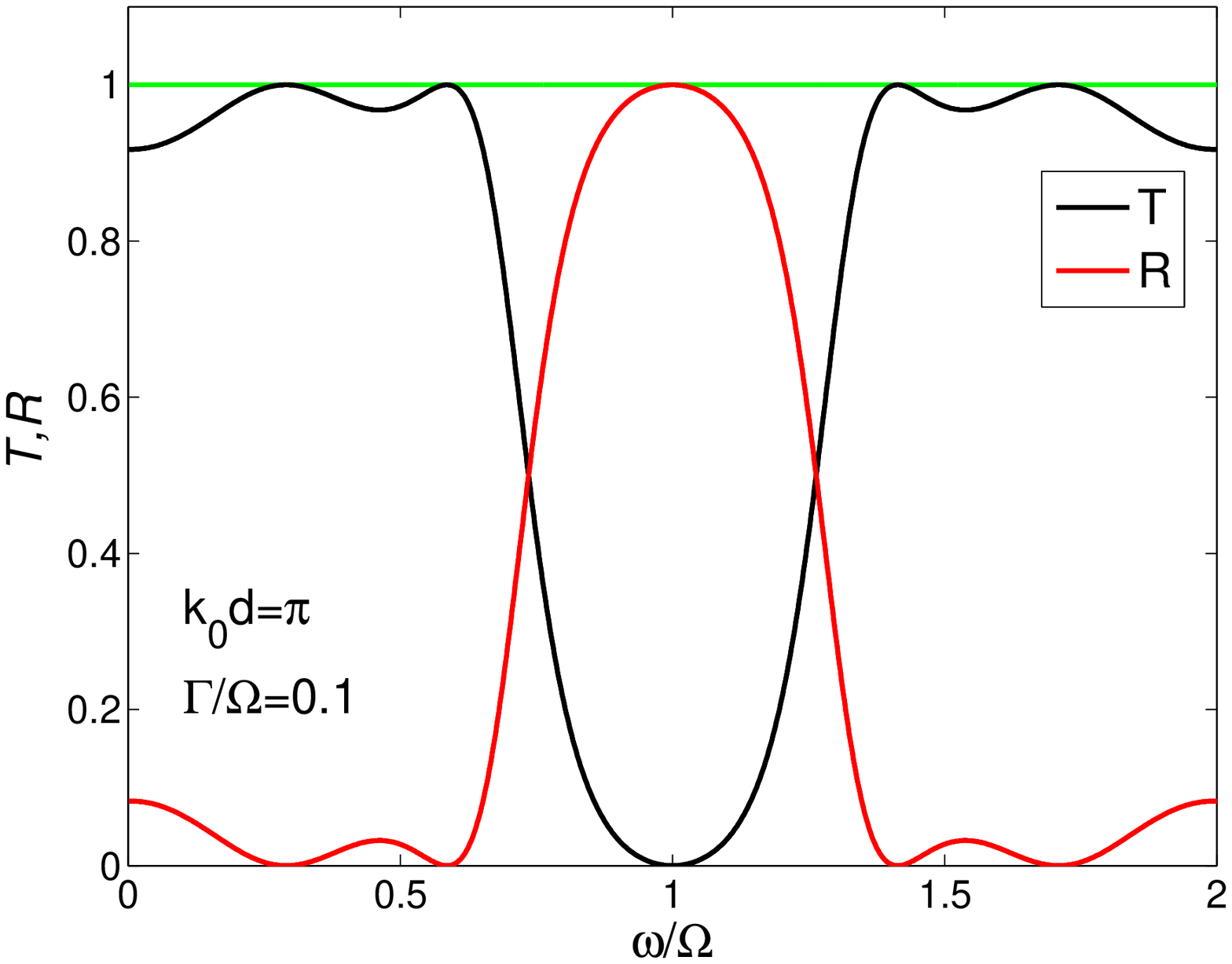}} c) \\
\end{minipage}
\hfill
\begin{minipage}[h]{0.47\linewidth}
\center{\includegraphics[width=1\linewidth]{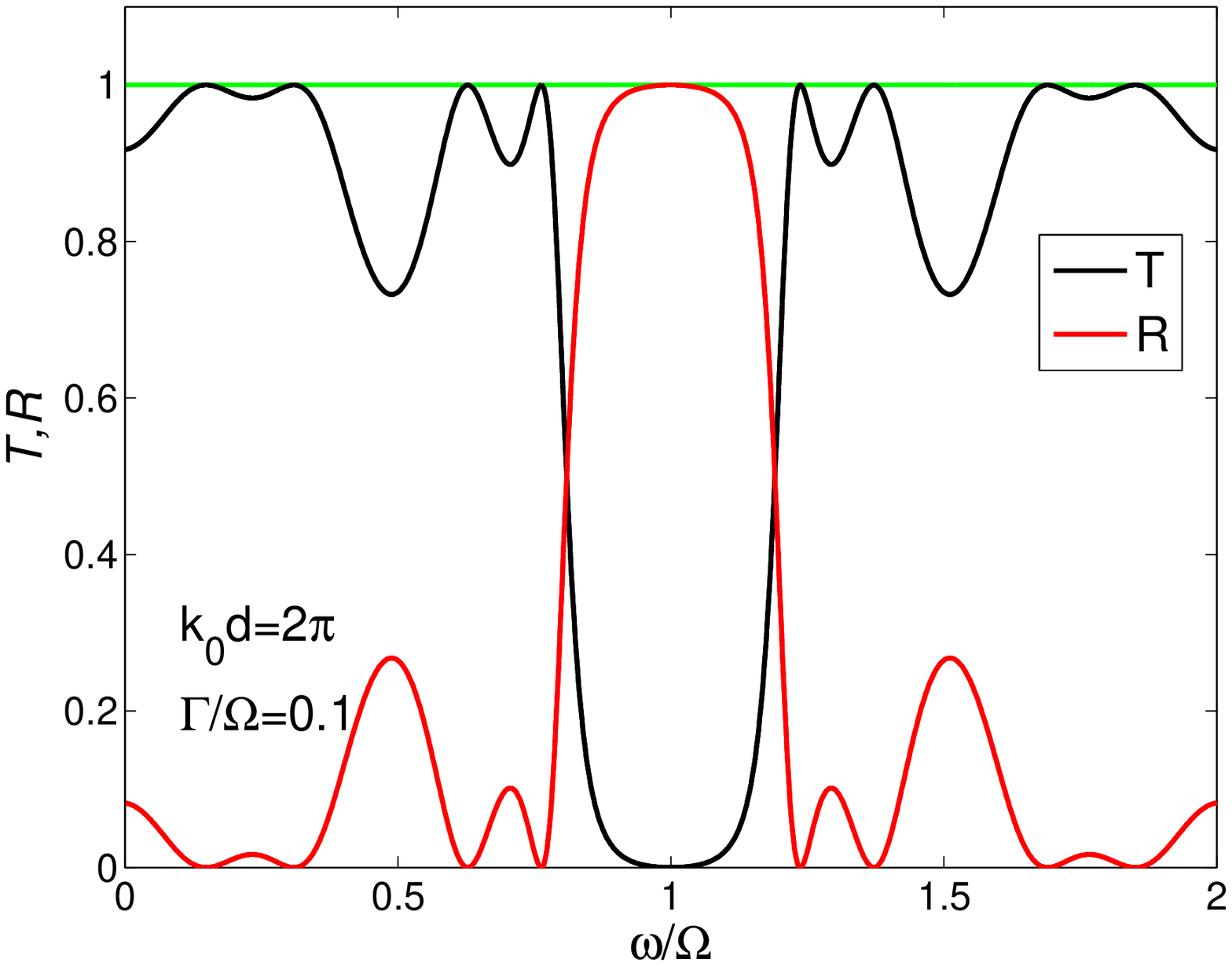}} d) \\
\end{minipage}
\caption{Color online. The dependence of transmission (black) and
reflection (red) amplitudes on the frequency of incident photon,
$\omega/\Omega$ for different values of $k_0d$ for three identical
qubits.} \label{3qbdata}
\end{figure}

\subsubsection{Photon mediated entanglement for three qubits}

Analogous to two qubit case the function of the qubit system
$\Psi_Q$ (\ref{psiQ1}) can be written as a linear superposition of
the three two-qubit states $\Psi_Q  = a\left| 1 \right\rangle  +
b\left| 2 \right\rangle+c\left| 3\right\rangle$, where, in
general, $a$, $b$, and $c$ depend on the physical frequency
$\omega$. For three identical qubits we obtain a general
expression which describes the frequency dependent entanglement of
three three-qubit states:
\begin{multline}\label{entangl_3}
    \Psi_Q=\frac{\lambda}{D_3^{id}(\omega)}
    \left([(\omega-\Omega+i\Gamma)^2e^{-ikd}+\Gamma^2e^{ikd}\right.\\
    -i\Gamma(\omega-\Omega+i\Gamma)e^{ikd}-i\Gamma(\omega-\Omega-i\Gamma)e^{3ikd}]|1\rangle\\
   \left.(\omega-\Omega)^2e^{ikd}|2\rangle+(\omega-\Omega)[\omega-\Omega+i\Gamma(1-e^{2ikd})]|3\rangle\right)
\end{multline}

In the long wavelength limit $kd<<1$ the maximally entangled
superradiant state which corresponds to a coherent symmetric
superposition of three three-qubit states is formed:

\begin{equation}\label{sym_entangl3}
    ({\Psi_Q) _S} = \frac{\lambda}{\omega-\Omega+3i\Gamma}
    \left( {\left| 1 \right\rangle  + \left| 2
\right\rangle }+|3\rangle \right)
\end{equation}

The transmission and reflection in this case are given by the
expressions (\ref{tidlw}) and (\ref{ridlw}). The resonance line of
superradiant state is directly given as the line of reflection
factor (\ref{ridlw}).

For arbitrary values of $kd$ maximally entangled states are formed
only for fixed values of the frequency $\omega$. For example, if
$kd\equiv\omega d/v_g=n\pi$ ($n=1,2,...$) we obtain from
(\ref{entangl_3}) the expression
\begin{equation}\label{entangl1}
    \Psi_Q=\frac{\lambda}{\omega_n-\Omega+3i\Gamma}\left[(-1)^n(|1\rangle+|2\rangle)+|3\rangle\right]
\end{equation}
where $\omega_n=n\pi v_g/d$.

For on resonant excitation ($\omega=\Omega$) and $k_0d\neq n\pi$,
where $k_0=\Omega/v_g$ we get from (\ref{entangl_3}) unentangled
state $\Psi_Q=i\left(\lambda /\Gamma\right)e^{ik_0d} |1\rangle$.
In this case we observe a full reflection with only the first
qubit being excited.

\subsubsection{Resonances in three- qubit system}

As in the two qubit case, the resonance frequencies in (\ref{3t})
and (\ref{3r}) are given by three equations
$\omega=\rm{Re}[\widetilde{\omega}_1(\omega)]$,
$\omega=\rm{Re}[\widetilde{\omega}_2(\omega)]$,$\omega=\rm{Re}[\widetilde{\omega}_3(\omega)]$,
where $\widetilde{\omega}_1, \widetilde{\omega}_2$ and
$\widetilde{\omega}_3$ are the roots of equation (\ref{roots}).

The denominator (\ref{Ft}) can then be written as
$D_3(\omega)=\left[\omega-\widetilde{\omega}_1(\omega)\right]$
$\left[\omega-\widetilde{\omega}_2(\omega))\right]$
$\left[\omega-\widetilde{\omega}_3(\omega))\right]$. Hence, the
resonance frequencies of the incident photon are given by the
roots of, in general, nonlinear equations
$\omega=\rm{Re}[\widetilde{\omega}_1(\omega)]$,
$\omega=\rm{Re}[\widetilde{\omega}_2(\omega)]$,
$\omega=\rm{Re}[\widetilde{\omega}_3(\omega)]$.

Below we consider three identical qubits. Similar to the case of
two qubits we define a spectral function by dividing the
transmission (\ref{tid}) by the factor
$[(\omega-\Omega)/\Omega]^3$:

\begin{equation}\label{SF_3}
    S(\omega)=\frac{\Omega^3}{D_3^{id}(\omega)}
\end{equation}
where $D_3^{id}(\omega)$ is given in (\ref{Fid}).

\begin{figure}[h]
  \includegraphics[width=0.8\columnwidth]{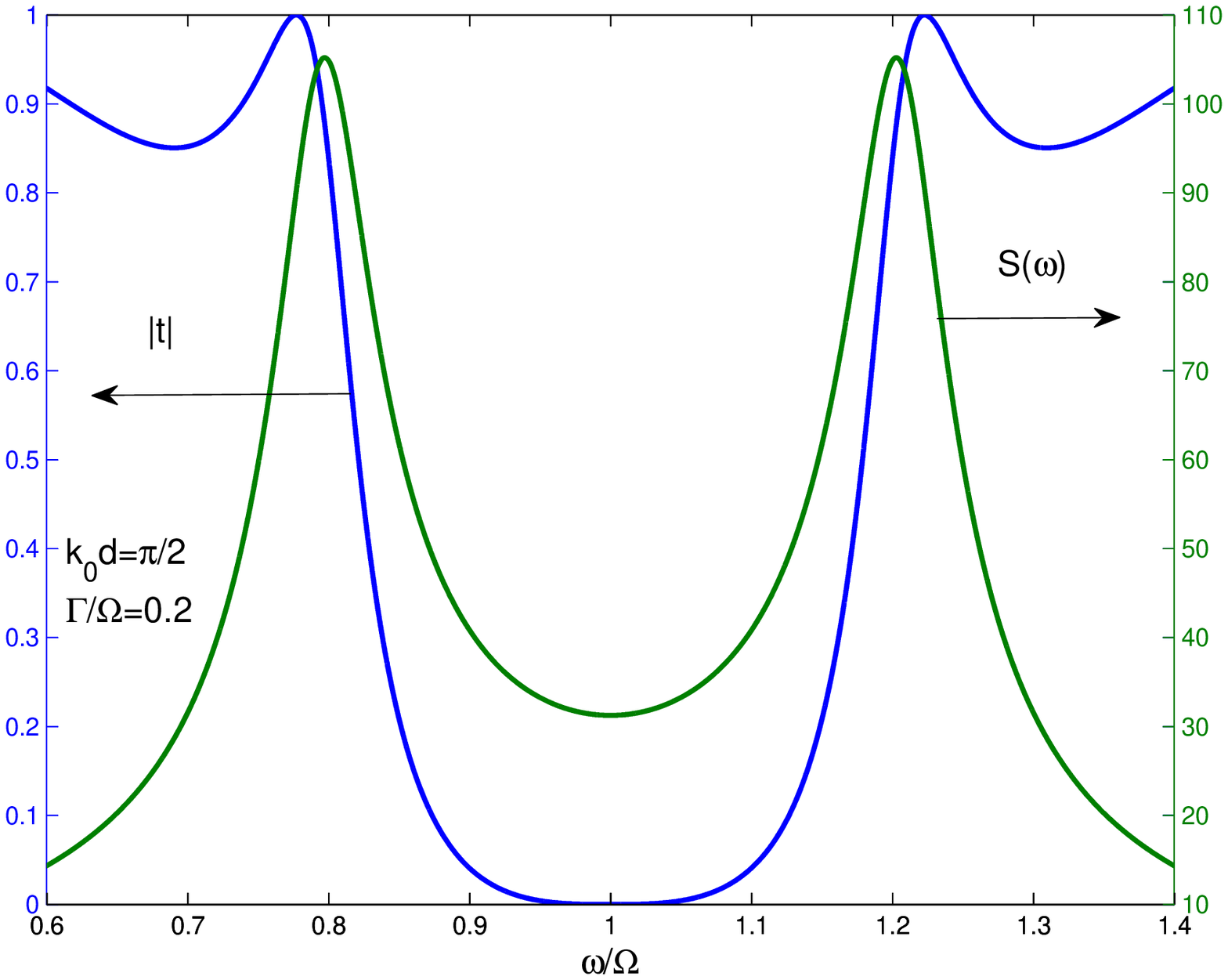}\\
  \caption{Color online. Frequency dependence of the transmission (left axis, blue line)
  and spectral function (right axis, green line)for three identical qubits.
  $k_0d=\pi/2, \Gamma/\Omega=0.2$.}\label{trans+spectr}
\end{figure}

Below  we show the plots of transmission and spectral function
$S(\omega)$ which exhibits peaks corresponding to the roots of the
equations $\omega=\rm{Re}[\widetilde{\omega}_i(\omega)]
(i=1,2,3)$, where $\widetilde{\omega}_1, \widetilde{\omega}_2$ and
$\widetilde{\omega}_3$ are the roots of the equation
(\ref{roots1}).  The resonance spectrum for $k_0d=\pi/2,
\Gamma/\Omega=0.2$ is shown at Fig.\ref{trans+spectr}. For this
case there are three resonances at the frequencies
$\omega/\Omega=0.8, 1.0, 1.2$ with corresponding half widths
$\widetilde{\Gamma}/\Omega=-0.046, -0.40,-0.046$.  However, only
two resonances which are closest to the frequency axis (with the
width $2\widetilde{\Gamma}/\Omega=0.08$),
 are visible in Fig.\ref{trans+spectr}.

\begin{figure}[h]
  \includegraphics[width=0.8\columnwidth]{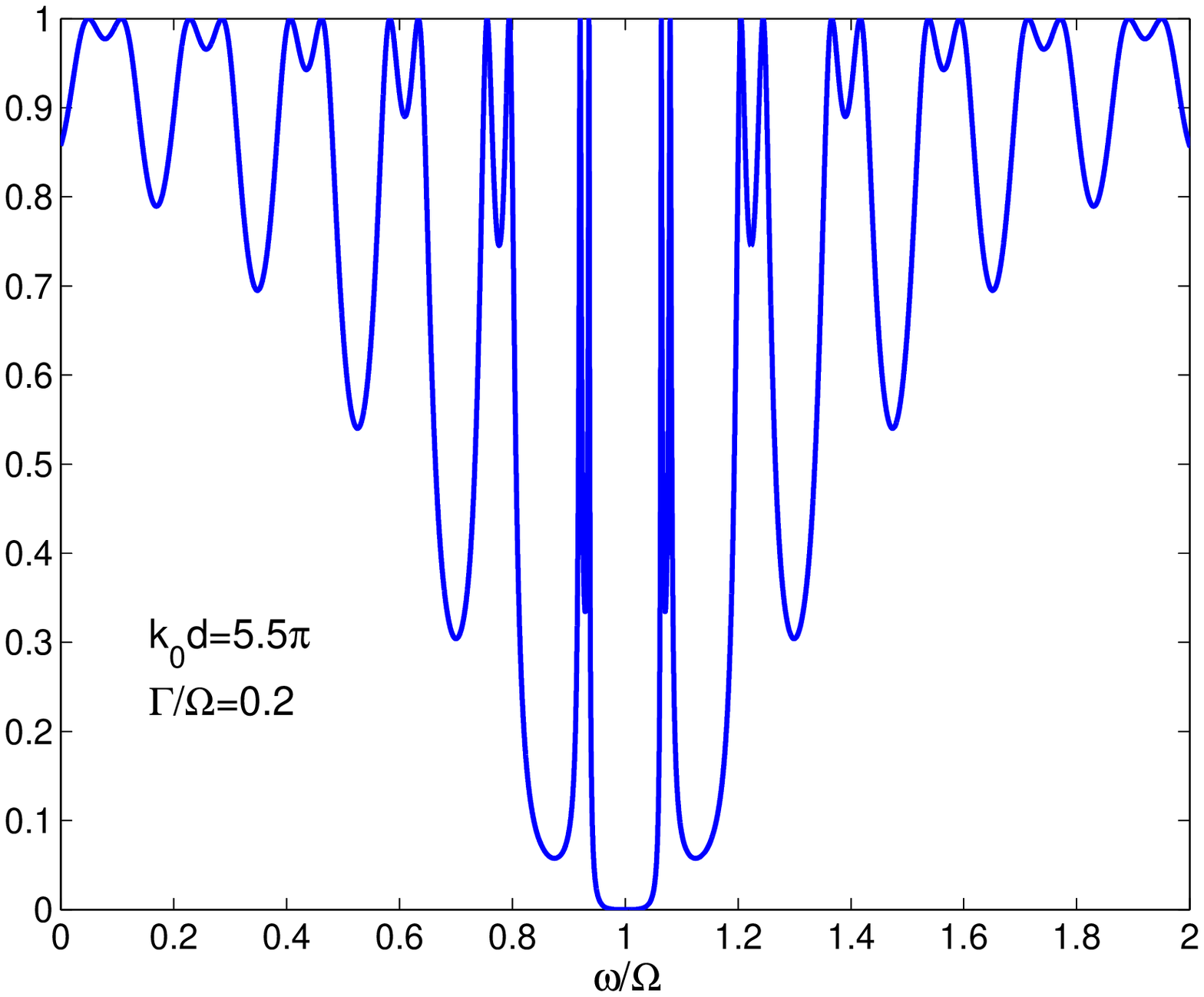}\\
  \caption{Color online. Transmission pattern for three identical qubits.
  $k_0d = 5.5\pi,  \Gamma/\Omega = 0.2$.}\label{trans3idQb}
\end{figure}

\begin{figure}[htp]
  \includegraphics[width=0.8\columnwidth]{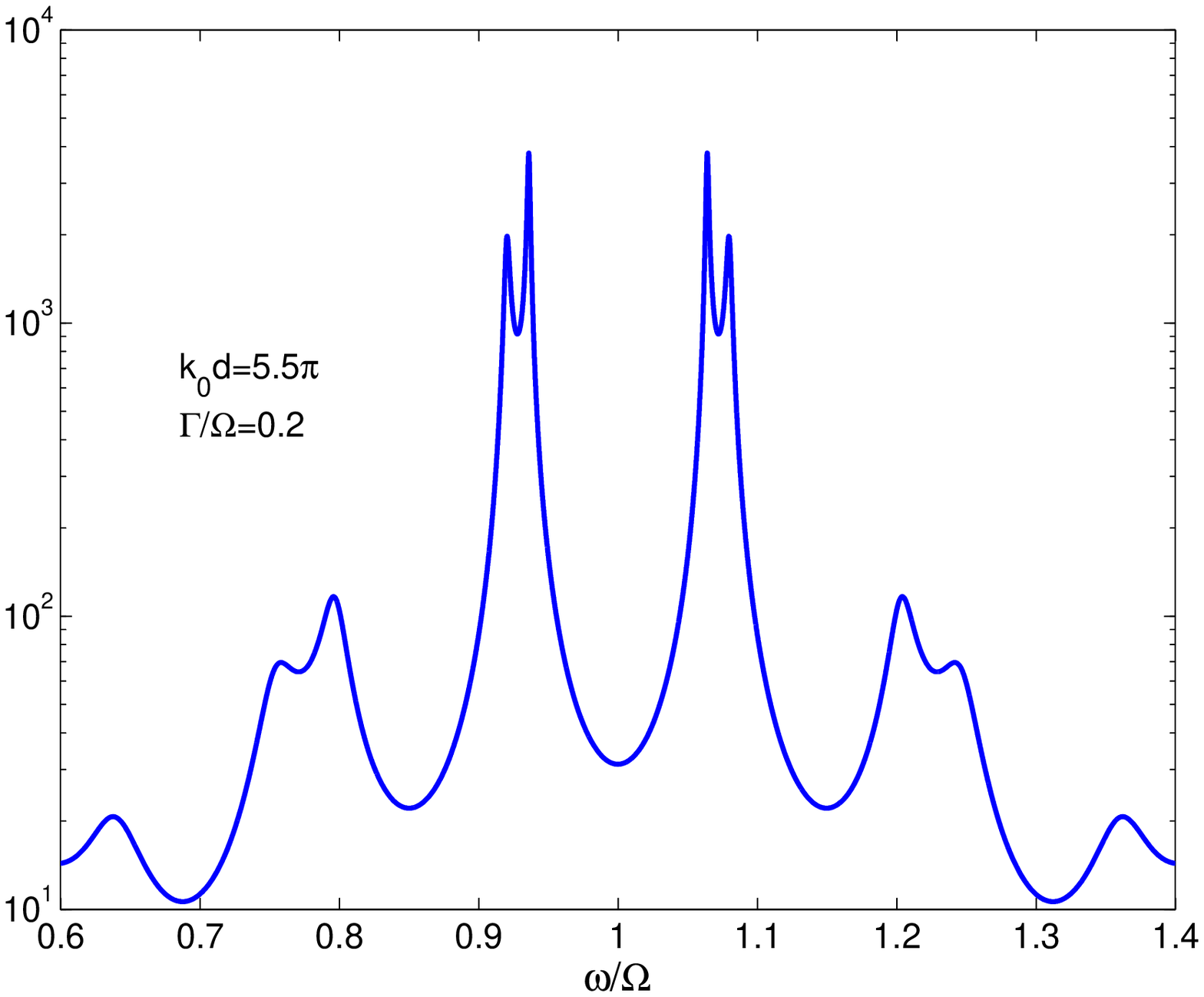}\\
  \caption{Color online. Frequency dependence of the spectral function for three identical
   qubits. $k_0d = 5.5\pi,  \Gamma/\Omega = 0.2$. The y-axis is in log scale.}\label{Spectr3idQb}
\end{figure}

With the increase of the inter qubit distance $d$, the number of
resonances are also increased, being in the vicinity of $\Omega$
for small $\Gamma$'s. The corresponding transmission pattern and
the resonance spectrum are shown in Fiq.\ref{trans3idQb} and
Fig.\ref{Spectr3idQb} for $k_0d=5.5\pi, \Gamma/\Omega=0.2$.

In this case in the range $\omega/\Omega=0.6\div1.4$ there are
$13$ resonance frequencies, ten of which are seen in
Fig.\ref{Spectr3idQb}. The two highest peaks have the half width
$\widetilde{\Gamma}/\Omega=3.5\times10^{-3}$ and
$\widetilde{\Gamma}/\Omega=1.63\times10^{-2}$, respectively. We
note that the widths of some resonances, which define the decay
rates of the superposition state (\ref{entangl_3}), are much
smaller than the width of individual qubit.

We note that, the points of the full transition which correspond
to the zeros of the reflection, do not coincide with the
frequencies of the resonances. While for some frequencies the
corresponding values can be close to each other, in general, many
reflection zeros are out of the range of resonance frequencies.

\subsubsection{Photon wave function for three identical qubits in a waveguide}

Photon wave function for three identical qubits is calculated from
(\ref{Ph4}) with the matrix $R_{mn}$ defined in (\ref{Rmnid}).

\begin{equation}\label{Psi3}
\begin{array}{l}
{\Psi _3}(x) = {e^{ikx}} - i\hbar \Gamma {e^{ik\left| {x - d} \right|}}\left[ {{e^{ikd}}{R_{11}} + {e^{ - ikd}}{R_{12}} + {R_{13}}}
\right]\\[3pt]
 - i\hbar \Gamma {e^{ik\left| {x + d} \right|}}\left[ {{e^{ikd}}{R_{12}} + {e^{ - ikd}}{R_{11}} + {R_{13}}}
 \right]\\[3pt]
 - i\hbar \Gamma {e^{ik\left| x \right|}}\left[ {{e^{ikd}}{R_{13}} + {e^{ - ikd}}{R_{13}} + {R_{33}}} \right]
\end{array}
\end{equation}
Outside the qubit array $x<-d$ and $x>d$ the wavefunction
(\ref{Psi3}) is similar to (\ref{1qb6}), where the transmission
$t$ and reflection $r$ are given in (\ref{tid}) and (\ref{rid}).
Inside the array we obtain
\begin{equation}\label{Psi_ins1}
{\Psi _3}(x) = \frac{{{{(\omega  - \Omega )}^2}}}{{D_3^{id}(\omega
)}}\left[ {{e^{ikx}}(\omega  - \Omega + i\Gamma ) - i\Gamma {e^{ -
ikx}}{e^{2ikd}}} \right]
\end{equation}
for $0<x<d$, and
\begin{equation}\label{Psi_ins2}
\begin{array}{l}
{\Psi _3}(x) = \frac{{(\omega  - \Omega )}}{{D_3^{id}(\omega
)}}\left[ {{e^{ikx}}{{(\omega  - \Omega  + i\Gamma )}^2} + {\Gamma
^2}{e^{2ikd}}} \right.\\[5pt] \left. { - i\Gamma {e^{ - ikx}}\left[
{(\omega  - \Omega  + i\Gamma ) + (\omega  - \Omega  - i\Gamma
){e^{2ikd}}} \right]} \right]
\end{array}
\end{equation}
for $-d<x<0$.

Similar to the two qubit case, here at the exact resonance
($\omega=\Omega$) the photon wavefunction does not penetrate in
the inter qubit region.

In the conclusion to this section we write the probability for the
particular qubit in the array to be in excited state. From
(\ref{prob1}) we obtain:
\begin{equation}\label{prob_1}
\begin{array}{l}
\left\langle {1|\Psi_Q } \right\rangle  = \frac{\lambda }{{\hbar
D_3^{id}(\omega )}}\left[ {{{\left( {\omega  - \Omega  + i\Gamma }
\right)}^2}{e^{ - ikd}}}
\right.\\[5pt]
\quad\quad - i\Gamma \left( {\omega  - \Omega  - i\Gamma }
\right){e^{3ikd}}\left. { - i\Gamma \left( {\omega  - \Omega  +
2i\Gamma } \right){e^{ikd}}} \right]
\end{array}
\end{equation}

\begin{equation}\label{prob_2}
\left\langle {2|\Psi_Q } \right\rangle  = \frac{\lambda }{\hbar
}\frac{{{{\left( {\omega  - \Omega }
\right)}^2}}}{{D_3^{id}(\omega )}}{e^{ikd}}
\end{equation}

\begin{equation}\label{prob_3}
\left\langle {3|\Psi_Q } \right\rangle  = \frac{\lambda }{\hbar
}\frac{{(\omega  - \Omega )}}{{D_3^{id}(\omega )}}\left[ {\omega -
\Omega  + i\Gamma \left( {1 - {e^{2ikd}}} \right)} \right]
\end{equation}
We see that at resonance the first qubit only is excited. This is
consistent with the above conclusion that at resonance photon does
not penetrate beyond the first qubit.

All qubit arrays considered above have a general property: if the
photon frequency is equal to the resonance frequency of any qubit
in the chain, the transmission signal is absent. We attributed
this property to the destructive interference between the input
wave and the wave which resulted from the virtual transitions
between the qubits and the photon field in the resonator. However,
it is not clear to what extent this property can be attributed to
uniform distribution of the qubit in the chain with the equal
distance between adjacent qubits. The simplest system where we can
check this property is the three qubit chain.  We made the
calculation of the transmission for three different qubits which
are positioned at the points $x_1=-d_1$, $x_2=+d_2$ and $x_3=0$,
respectively, with unequal distance between adjacent qubits. It
turned out that in this case the transmission is similar to
(\ref{3t}) with the just the same numerator, but different
denominator, which is given in the Appendix. Hence, we may assume
that for nonuniform qubit array the transmission is also zero if
the input photon is at resonance with any qubit in the chain.

\subsubsection{The manipulation of the photon transmission
in three qubit chain}

\begin{figure}[h]
  \includegraphics[width=0.8\columnwidth]{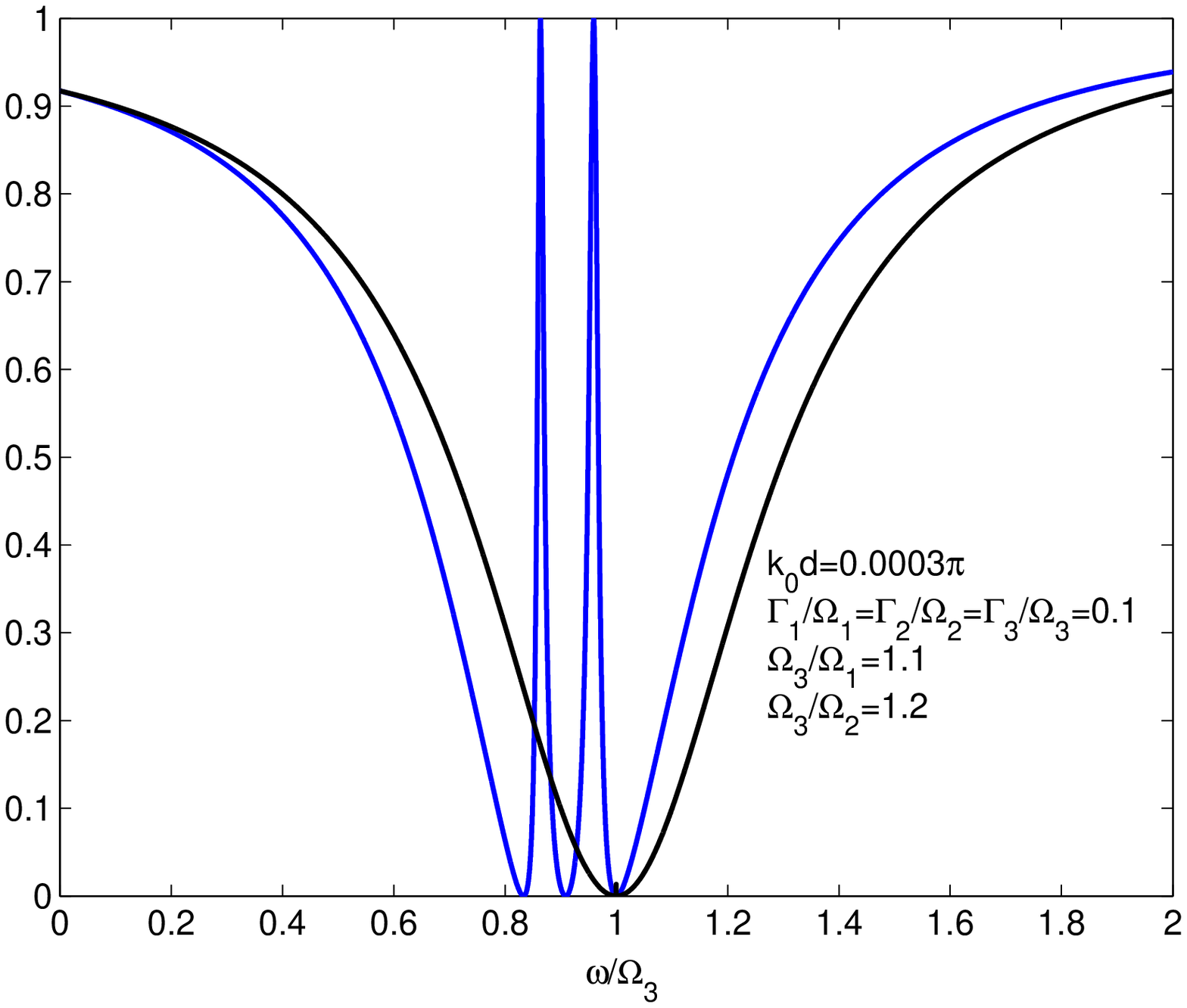}\\
  \caption{Color online. Transmission for three non-identical qubits (blue
  line). For comparison the transmission for identical qubits is also shown (black line)}\label{trans3_not_id_Qb}
\end{figure}

Most of artificial atoms, which are used as qubits, can be
addressed individually, so that every qubit frequency, $\Omega_i$,
in the chain can be tuned from external source. Here we show how
this property can be used to manipulate the photon transmission
through a waveguide. As an example we consider three non-identical
qubits, which are positioned at the points $x_1=-d$, $x_2=+d$ and
$x_3=0$, respectively, with a distance $d$ between adjacent
qubits. The frequencies $\Omega_1$ and $\Omega_2$ correspond to
qubits at the points $x_1, x_2$, respectively, while $\Omega_3$
corresponds to central qubit at the point $x_3$. In general, if
all three frequencies are different we obtain for long wavelength
($kd<<1$) transmission the plot shown in
Fig.\ref{trans3_not_id_Qb}, where for comparison the transmission
for identical qubits are also shown. The qubit frequencies are
directly given by zeros of the transmission, while two narrow
nearby peaks, which lie between qubit frequencies correspond to
the full transmission.

\begin{figure}[h]
  \includegraphics[width=0.8\columnwidth]{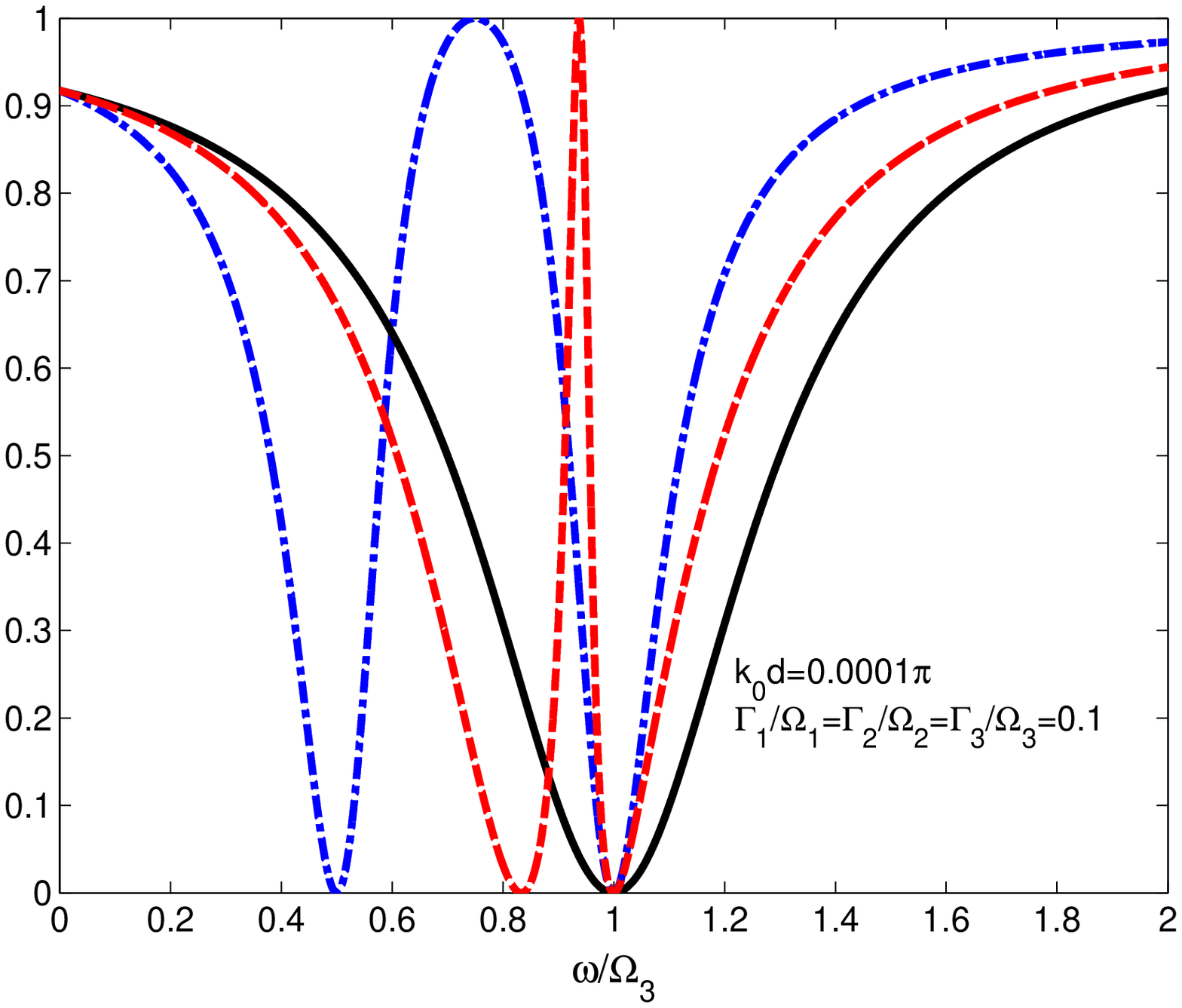}\\
  \caption{Color online. Transmission for three qubits, two of which are identical. Dashed (red)  line:
 $\Omega_3/\Omega_1=\Omega_3/\Omega_2=1.2$. Dash-dotted (blue) line:
  $\Omega_3/\Omega_1=\Omega_3/\Omega_2=2$. For comparison the transmission for identical
  qubits is also shown by solid (black) line.}\label{trans_3_non_id_Qb_1}
\end{figure}

Suppose now that the frequencies of the left and right qubit are
equal ($\Omega_1=\Omega_2\equiv\Omega$) and we manipulate the
frequency of the central qubit, $\Omega_3$. Then we obtain the
picture like the one shown in Fig.\ref{trans_3_non_id_Qb_1}.  By
changing the qubit frequency $\Omega_3$ we may move the frequency
of the transmission resonance between $\Omega$ and $\Omega_3$ and
manipulate its width. When $\Omega_3$ becomes equal to $\Omega$
the transmission signal disappear.

\subsection{$N$ qubits in a waveguide}
In principle the transmission and reflection for any number of
qubits can be found from general expressions (\ref{tgen}) and
(\ref{rgen}). For a chain of $N$ identical homogeneously
distributed two level atoms the analytical expression for the 1D
transmission was found in \cite{Tsoi08}. While in general case it
is not easy to find analytical solutions for $N$ qubits,
nevertheless, from previous calculations of the transmission for
one, (\ref{1qb4}), two, (\ref{2qbt}), and three (\ref{3t}) qubits,
we may guess the general structure of the transmission for $N$
qubits in a waveguide:

\begin{equation}\label{tN}
{t_N} = \frac{{\prod\limits_{n = 1}^N {(\omega  - {\Omega _n})}
}}{{{D_N}(\omega )}}
\end{equation}
where
\begin{equation}\label{DN}
{D_N}(\omega ) = \det \left( {\omega  - \frac{1}{2}\sum\limits_{i
= 1}^N {{\Omega _i}}  - {H_{eff}}/\hbar } \right)
\end{equation}

As for the spectral properties of the effective Hamiltonian, we
can assert that the secular equation $\rm{det}$$(E-H_{eff})$ has
$N$ poles in the low half plain of the complex energy. For
identical qubits in the long wavelength limit there are $N-1$
stable states with $\widetilde{\omega}=\Omega$ and one resonance
$\widetilde{\omega}=\Omega-i N\Gamma$ which absorbs all the widths
of individual qubits:

\begin{equation}\label{tidNlw}
t_N = \frac{{\omega  - \Omega }}{{\omega  - \Omega + iN\Gamma}}
\end{equation}

\begin{equation}\label{ridNlw}
r_N =  - \frac{{iN \Gamma }}{{\omega  - \Omega
 + iN\Gamma }}
\end{equation}

\section{conclusion}
In this paper we develop a new technique for the investigation of
the photon transport through multiple qubit array in a 1D
waveguide. The technique is based on the projection operators
formalism and non Hermitian approach, which is known to be a
successful tool in some fields of nuclear physics and condensed
matter. We considered in detail the one photon transport for two
and three qubits in a waveguide, and made some conclusions for $N$
qubit case. We showed that the interaction of qubits with a photon
field results in the frequency dependent superposition of the
qubit states. We investigated in detail the resonance spectra for
two and three qubits and showed that in non Markovian case
($kd>>1$) the resonance widths, which define the decay rates of
the superposition state, can be much smaller than the decay width
of individual qubit.

We also showed that in the long wavelength limit for uniformly
distributed array of identical qubits  a coherent superradiance
state is formed with the width being equal to the sum of the
widths of spontaneous transitions of \emph{N} individual qubits.
Within the framework of our method it is not difficult to account
for the decay of the qubit states to the modes other than the
waveguide continuum. It can be done by simply adding an imaginary
term in the qubit energy level,
$\Omega_n\rightarrow\Omega_n-i\Gamma_n'$.

The approach developed in the paper can be easily generalized to
include the exchange interaction $H_J$ between nearest neighbor
qubits:

\begin{equation}\label{HJ}
    {H_J} = \hbar\sum\limits_{i = 1}^N {{J_i}\left( {\sigma _ + ^i\sigma _
- ^{i + 1} + \sigma _ + ^{i + 1}\sigma _ - ^i} \right)}
\end{equation}
For this case it is necessary to change only the matrix of
effective Hamiltonian (\ref{Hij2}):
\begin{multline}\label{H_ex}
    \left\langle {m} \right|{H_{eff}}\left| {n} \right\rangle  =
\varepsilon_m\delta_{m,n}+\hbar J_{n-1}\delta_{m,n-1}+\hbar J_n\delta_{m,n+1}\\
   -i\hbar(\Gamma_m\Gamma_n)^{1/2}e^{ik|d_{mn}|}
\end{multline}

 The results obtained in the paper are of general
nature and can be applied to any type of qubits. The specific
properties of the qubit are encoded in only two parameters: the
qubit energy $\Omega$ and the rate of spontaneous emissin
$\Gamma$. For example, for a superconducting flux qubit
$\,\Omega=\sqrt{\varepsilon^2+\Delta^2}$ where $\varepsilon$ is an
external parameter which by virtue of external magnetic flux,
$\Phi_X$ controls the gap between ground and excited states
\cite{Wal00}, and the quantity $\Delta$ is the qubit's gap at the
degeneracy point ($\varepsilon=0$). The rate of spontaneous
emission $\Gamma=g\Delta/\Omega$ \cite{Om10}, where $g$ is the
qubit- waveguide coupling.

\begin{acknowledgments}
Ya. S. G. thanks A. Satanin, V. Zelevinsky, A. Fedorov, and
Yao-Lung L. Fang for fruitful discussions. The work is supported
by the Ministry of Education and Science of Russian Federation
under the project 3.338.2014/K.
\end{acknowledgments}

\appendix*
\section{}
\subsection{The calculation of integral
$J(x_m,x_n)$(\ref{Jij})}\label{ap1}

In (\ref{Jij}) the energies $E_q$ and $E_k$ are the energies of
incident photon $|q\rangle$ ($|k\rangle$) plus the energy of $N$
qubits in the ground state. Hence,
$E_q-E_k=\hbar(\omega_q-\omega_k)$.  For (\ref{Jij}) we,
therefore, have
          \begin{equation}\label{a6}
    J(x_m,x_n) =\frac{1}{\hbar}\int\limits_{ - \infty }^{ + \infty }
    {dk\frac{{{e^{ikd_{mn}}}}}{{{\omega _q} - {\omega _k} + i\varepsilon }}}
\end{equation}

The main contribution to this integral comes from the region
${\omega _k} \approx {\omega _q}$. Since  $\omega_k$ is the even
function of $k$, the poles of the integrand (\ref{a6}) in the $k$
plane are located near the points $k\approx\pm q$. For an
arbitrary frequency $\omega_q$ that is away from the cutoff of the
dispersion, with the corresponding wave vector $\pm q$ , we
approximate $\omega_k$ around $+q$ and $-q$ as
 \begin{equation}\label{a7}
    {\omega _k} \approx {\omega _q} + (k - q){\left. {\frac{{d{\omega _k}}}{{dk}}}
    \right|_{k =  + q}} = {\omega _q} + (k - q){{\mathop{\rm v}\nolimits} _g}
\end{equation}
   \begin{equation}\label{a8}
    {\omega _k} \approx {\omega _q} + (k + q){\left. {\frac{{d{\omega _k}}}{{dk}}}
    \right|_{k =  - q}} = {\omega _q} - (k + q){{\mathop{\rm v}\nolimits} _g}
\end{equation}
Near the poles the denominator in (\ref{a6}) takes the form:
   \begin{equation}\label{a9}
    - {{\mathop{\rm v}\nolimits} _g}(k - q) + i\varepsilon
\end{equation}
\begin{equation}\label{a10}
    {{\mathop{\rm v}\nolimits} _g}(k + q) + i\varepsilon
\end{equation}
Therefore, one pole is located in the upper half of the $k$ plane,
$k=q+i\varepsilon$, the other pole is located in the lower half of
the $k$ plane, $k=-q-i\varepsilon$. For positive $d_{mn}$, when
calculating the integral (\ref{a6}) we must close the path in the
upper plane. For negative $d_{mn}$ the path should be closed in
lower plane. Thus, we obtain:

\begin{equation}\label{A0}
J(x_m,x_n) =  - \frac{{2\pi i}}{\hbar v_g}{e^{ik\left|
d_{mn}\right|}}
\end{equation}

\subsection{Calculation of the $R$ matrix for three qubits in a
waveguide}\label{ap2}

The matrix $R_{m,n}$, ($m,n=1,2,3$)  is calculated as the inverse
of the matrix $(E-H_{eff})_{m,n}$, the elements of which can be
found from (\ref{Hmtr}). The matrix $R_{m,n}$ is symmetric so that
$R_{12}=R_{21},R_{13}=R_{31}, R_{23}=R_{32} $. Direct calculations
yield for $R_{m,n}$ the following result:
\begin{equation}\label{Rmn3}
\begin{array}{l}
{D_3}(E){R_{11}} = \left( {E - {\varepsilon _2} + i\hbar {\Gamma
_2}} \right) \left( {E - {\varepsilon _3} + i\hbar {\Gamma _3}}
\right)\\
 + {\hbar ^2}{\Gamma _2}{\Gamma
_3}{e^{2ikd}}\\
{D_3}(E){R_{22}} = \left( {E - {\varepsilon _1} + i\hbar {\Gamma
_2}} \right) \left( {E - {\varepsilon _3} + i\hbar {\Gamma _3}}
\right) \\
+ {\hbar ^2}{\Gamma _1}{\Gamma
_3}{e^{2ikd}}\\
{D_3}(E){R_{33}} = \left( {E - {\varepsilon _2} + i\hbar {\Gamma
_2}} \right) \left( {E - {\varepsilon _1} + i\hbar {\Gamma _3}}
\right) \\
+ {\hbar ^2}{\Gamma _1}{\Gamma
_2}{e^{4ikd}}\\
{D_3}(E){R_{12}} =  - \left( {E - {\varepsilon _3} + i\hbar
{\Gamma _3}} \right)i\hbar \sqrt {{\Gamma _1}{\Gamma _2}}
{e^{2ikd}}\\
{- {\hbar ^2}\sqrt {{\Gamma _1}{\Gamma _2}} {\Gamma
_3}{e^{2ikd}}}\\
{D_3}(E){R_{13}} =  - \left( {E - {\varepsilon _2} + i\hbar
{\Gamma _2}} \right)i\hbar \sqrt {{\Gamma _1}{\Gamma _3}}
{e^{ikd}}\\
 - {\hbar ^2}\sqrt {{\Gamma _1}{\Gamma _3}} {\Gamma
 _2}{e^{3ikd}}\\
{D_3}(E){R_{23}} =  - \left( {E - {\varepsilon _1} + i\hbar
{\Gamma _1}} \right)i\hbar \sqrt {{\Gamma _2}{\Gamma _3}}
{e^{ikd}} \\- {\hbar ^2}\sqrt {{\Gamma _2}{\Gamma _3}} {\Gamma
_1}{e^{3ikd}}
\end{array}
\end{equation}

where $D_3(E)=\rm{det}\left(\emph{E}-\emph{H}_{eff}\right)_{mn}$:

\begin{equation}\label{D3}
\begin{array}{l}
{D_3}(E) = \left( {E - {\varepsilon _1} + i\hbar {\Gamma _1}}
\right)\left( {E - {\varepsilon _2} + i\hbar {\Gamma _2}}
\right)\left( {E - {\varepsilon _3} + i\hbar {\Gamma _3}}
\right)\\[3pt]
 + \left( {E - {\varepsilon _1} + i\hbar {\Gamma _1}} \right){\hbar ^2}{\Gamma _2}{\Gamma _3}{e^{2ikd}} +
 \\[3pt]
\left( {E - {\varepsilon _2} + i\hbar {\Gamma _2}} \right){\hbar
^2}{\Gamma _1}{\Gamma _3}{e^{2ikd}} + \left( {E - {\varepsilon _3}
- i\hbar {\Gamma _3}} \right){\hbar ^2}{\Gamma _1}{\Gamma
_2}{e^{4ikd}}
\end{array}
\end{equation}
and the quantities $\varepsilon_i (i=1,2,3)$ are defined in
(\ref{Hij1}).

At the end of this subsection we write down from (\ref{Rmn3}) the
matrix $R_{m,n}$ for three identical qubits.
\begin{equation}\label{Rmnid}
\begin{array}{l}
\hbar D_3^{id}(\omega ){R_{11}} = {\left( {\omega  - \Omega  + i\Gamma } \right)^2} + {\Gamma ^2}{e^{2ikd}}\\
\hbar D_3^{id}(\omega ){R_{33}} = {\left( {\omega  - \Omega  + i\Gamma } \right)^2} + {\Gamma ^2}{e^{4ikd}}\\
\hbar D_3^{id}(\omega ){R_{12}} =  - i\Gamma \left( {\omega  - \Omega } \right){e^{2ikd}}\\
\hbar D_3^{id}(\omega ){R_{13}} =  - i\Gamma \left( {\omega  -
\Omega
+i\Gamma } \right){e^{ikd}} - {\Gamma ^2}{e^{3ikd}}\\
R_{22}=R_{11}, R_{23}=R_{13}=R_{31}=R_{32}, R_{21}=R_{12}
\end{array}
\end{equation}
where
\begin{equation}\label{D3omega}
\begin{array}{l}
D_3^{id}(\omega ) = {\left( {\omega  - \Omega  + i\Gamma }
\right)^3} + 2{\Gamma ^2}\left( {\omega  - \Omega  + i\Gamma }
\right){e^{2ikd}}\\[3pt]
 + {\Gamma ^2}\left( {\omega  - \Omega  - i\Gamma } \right){e^{4ikd}}
\end{array}
\end{equation}\\[10pt]

\subsection{The transmission for three qubit chain with unequal
distance between each other}

Here we consider three different qubits which are positioned at
the points $x_1=-d_1$, $x_2=+d_2$ and $x_3=0$, respectively, with
$d_1\neq d_2$. The calculations yields the result:
\begin{equation}\label{app3t}
t_3= \frac{{(\omega  - {\Omega _1})(\omega  - {\Omega _2})(\omega
- {\Omega _3})}}{{F\left( \omega  \right)}}
\end{equation}

where
\begin{equation}
\begin{array}{l}
F(\omega ) = {U_1}{U_2}{U_3} + i{\Gamma _1}{\Gamma _2}{\Gamma _3}\left( {{e^{2ik{d_1}}} + {e^{2ik{d_2}}} - {e^{2ik({d_1} + {d_2})}} - 1}
\right)\\[3pt]
 + {U_3}{\Gamma _1}{\Gamma _2}\left( {{e^{2ik{d_1} + 2ik{d_2}}} - 1} \right) + {U_1}{\Gamma _2}{\Gamma _3}\left( {{e^{2ik{d_2}}} - 1}
 \right)\\[3pt]
 + {U_2}{\Gamma _1}{\Gamma _3}\left( {{e^{2ik{d_1}}} - 1} \right) + i{U_1}{U_2}{\Gamma _3} + i{U_1}{U_3}{\Gamma _2} + i{U_2}{U_3}\Gamma
\end{array}
\end{equation}
\[U_1=\omega-\Omega_1;\, U_2=\omega-\Omega_2;\,U_3=\omega-\Omega_3\]

The equation (\ref{app3t}) is similar to (\ref{3t}) with the same
numerator but different denominator.

\end{document}